\documentclass[pdflatex,sn-mathphys-num]{sn-jnl}

\usepackage{graphicx}%
\usepackage{multirow}%
\usepackage{amsmath,amssymb,amsfonts,bm}%
\usepackage{amsthm}%
\usepackage{mathrsfs}%
\usepackage[title]{appendix}%
\usepackage{xcolor}%
\usepackage{textcomp}%
\usepackage{manyfoot}%
\usepackage{booktabs}%
\usepackage{algorithm}%
\usepackage{algorithmicx}%
\usepackage{algpseudocode}%
\usepackage{listings}%
\usepackage{soul}

\theoremstyle{thmstyleone}%
\theoremstyle{thmstyletwo}%

\theoremstyle{thmstylethree}%

\begin{document}

\title[Article Title]{Realization of a universal topological waveguide by tuning adiabatic geometry}



\author*[1]{\fnm{Keita} \sur{Funayama}}\email{funayama@mosk.tytlabs.co.jp}

\author[1]{\fnm{Jotaro J.} \sur{Nakane}}

\author[2]{\fnm{Ai} \sur{Yamakage}}

\affil*[1]{\orgname{Toyota Central R{\&}D Labs., Inc.}, \orgaddress{\street{41-1, Yokomichi}, \city{Nagakute}, \postcode{4801192}, \state{Aichi}, \country{Japan}}}

\affil[2]{\orgdiv{Department of Physics}, \orgname{Nagoya University}, \orgaddress{\street{Furo-cho, Chikusa-ku}, \city{Nagoya}, \postcode{4648602}, \state{Aichi}, \country{Japan}}}


\abstract{
    Quantum valley Hall-based topological phases have been attracting attention across diverse fields as a robust platform for wave guidance due to their high compatibility with engineering frameworks.
    Combining three representative boundary types enables topological waveguides with flexible designability and enhanced functionality.
    However, one of the three, namely the armchair boundary, has long been limited by inter-valley scattering, resulting in weak topological protection and severely restricting its use in practical devices.
    This long-standing constraint is a major barrier to realizing broadly applicable topological waveguide systems.
    Here, to address this challenge toward a broadly applicable design framework for topological waveguides, we experimentally demonstrate that topological adiabatic geometry implemented in a micro electromechanical system suppresses valley mixing.
    We found that the adiabaticity enhances immunity to defects and increases the transmission efficiency of the armchair boundary.
    As the adiabaticity increases, topological protection is recovered over an increasingly broad portion of the bulk band gap, extending from low to high frequencies. 
    Furthermore, we show that the recovery of protection in the adiabatic armchair boundary enables waves to propagate through 90$^{\circ}$ and 150$^{\circ}$-bent waveguides by coupling with other interface geometries.
    Suppressing valley mixing via adiabaticity paves the way for a universal design framework for topological waveguides and for restoring robust topological characteristics across a wide range of wave phenomena. 
}




\maketitle

\section{Introduction}\label{sec1}

Topological physics has emerged as a powerful framework for controlling classical waves\cite{W-Yan2021,Y-Wang2025,Zheng2022,Miniaci2021,Lin2021,YuS2018,Y-Li2018,WangZ2009,ChaJ2018,Zilberberg2018,Serra2019,ZhaoH2019}, quantum waves\cite{Hashemi2025,HeL2024,WangM2019,Sabyasachi2018,Balco2018}, and diffusion phenomena\cite{HuH2022,QiM2022,WuH2023,FunayamaAPL2024,Funayama2023,LiuZ2024,Fukui2023,Yoshida2021,LiY2019,Nakane2025}.
In particular, topological phases arising from spatial-inversion-symmetry breaking with asymmetric sub-lattices, i.e., the quantum valley Hall (QVH) model\cite{DongJ2017,LuJ2017}, allow us to simplify device design, attracting a wide range of fields from fundamental physics to applied engineering.
Indeed, the QVH-based systems have successfully demonstrated unidirectional propagation of optical\cite{Jia2025,XueH2021,WuX2017,HeX2019,Hironobu2020,Kumar2024,Iwamoto2021,Zhihao2022,Ozawa2019,Muis2025}, acoustic\cite{ZhangX2018,WangJ2022,ZhangZ2021,ZhouY2023,YangZ2015,MaG2019}, and elastic waves\cite{MaJ2021,Funayama2025,YanM2018,Darabi2020} under outstanding robustness.

In QVH-based systems, the most widely used boundary/interface geometries are zigzag, bridge, and armchair.
By combining these three boundary types, the QVH systems provide a versatile design space for topological waveguides with flexible routing and enhanced functionality.
Out of the three, the zigzag and bridge boundaries exhibit the topological protection around K or K$^{\prime}$ points (valley) in momentum space.
On the other hand, in the case of the armchair boundary, the topological protection is broken by inter-valley scattering via a mixing of wave functions.
Thus, in conventional studies, the armchair boundary has been avoided to utilize as topological waveguides\cite{Funayama202412,WangM2019,MonikaDevi2021,WongS2020}, and the realization of universal waveguides has been largely abandoned.

The adiabatic approach has great potential to overcome the inter-valley mixing.
The adiabaticity around topological interfaces is introduced by smoothly varying the effective mass term corresponding to an asymmetric factor between sub-lattices.
Such an approach has been reported to untangle the pseudo-spin mixing which leads to unwanted backscattering and a degradation of topological protection\cite{Vakulenko2023}.
Applying the adiabatic scheme to the QVH-based armchair boundary, even the abandoned interface geometry, armchair, can be salvaged and work as a robust topological waveguide.
In QVH-based systems, the availability of all three boundaries significantly enhances the functionality and design flexibility of topological waveguides.
Previous studies have used adiabatic geometries primarily to improve the quality factor and propagation of QVH waveguides based on zigzag and bridge boundaries, which already exhibit relatively strong topological protection. 
However, applying adiabatic design to armchair boundaries, where inter-valley scattering severely weakens protection, has remained largely unexplored. 

Here, we numerically and experimentally demonstrate that the QVH-based adiabatic geometry untangles the mixing of wavefunctions between K and K$^{\prime}$ points in momentum space for the armchair boundary, with theoretical verification.
The adiabaticity-induced topological nature in the armchair boundary broadens the bandwidth enabling topologically protected wave propagation.
Our systematic approach revealed that topological protection recovers within the bulk band gap from low to high frequencies as the adiabatic factor increases.
Further, the combination of the adiabatic armchair boundary with other boundaries enables the design of the topological waveguides bent by 150$^{\circ}$ and 90$^{\circ}$ as well as 120$^{\circ}$ with remarkably high transfer efficiency within the entire bulk band gap.
We believe that our fundamental understanding of the untanglement of valley-mixing contributes to the high functionality of all kinds of wave and diffusion-based topological systems.

\section{Results}
\subsection{Untanglement of valley mixing by adiabatic domain wall}
\begin{figure}[t!]
    \centering
    \includegraphics[width=1.0\textwidth]{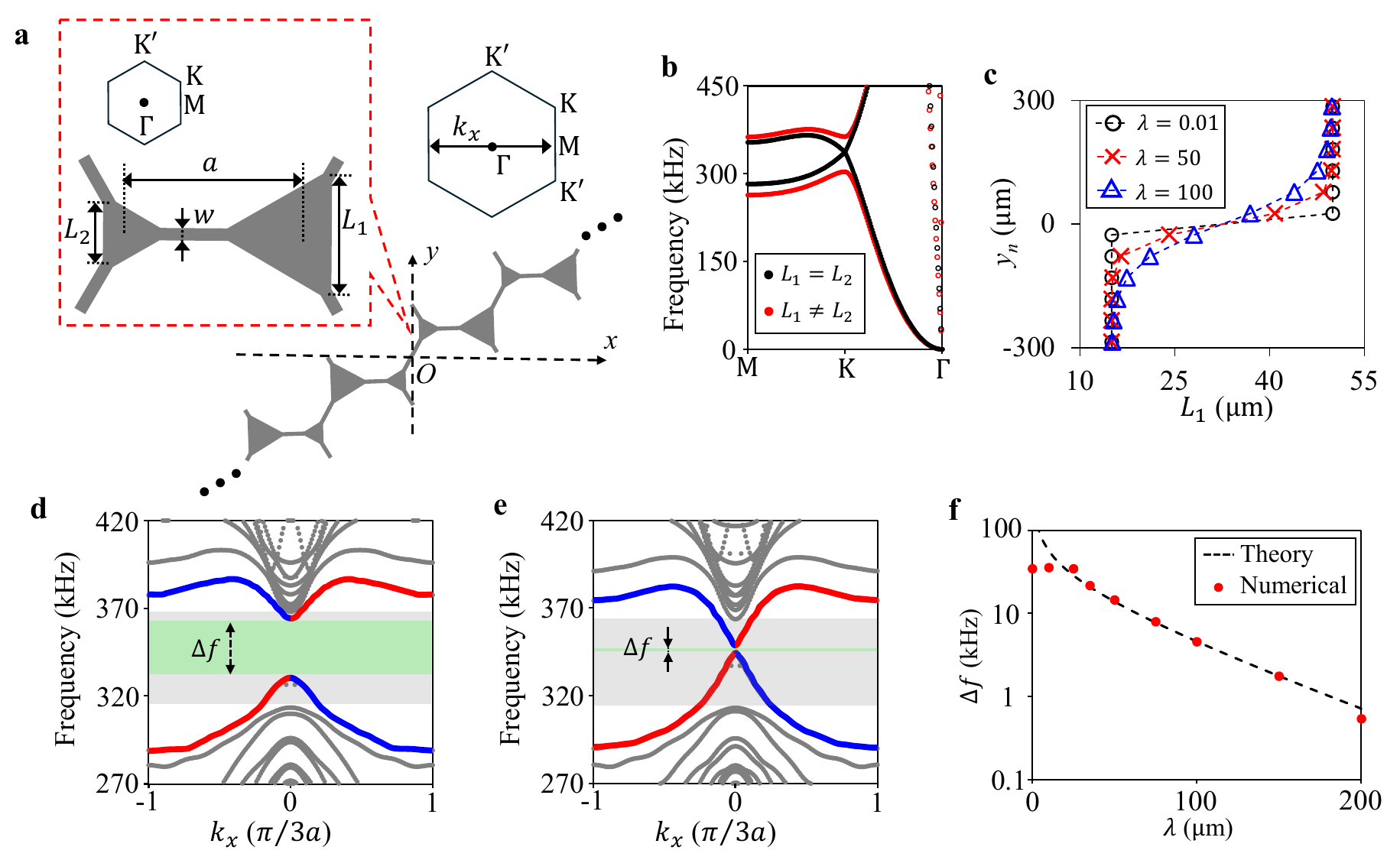}
    \caption{
        \textbf{Tunable adiabatic armchair boundaries and dispersion diagrams.}
        \textbf{a} Schematics of quantum valley-Hall (QVH)-based unit cell and supercell having an armchair boundary.
        In a numerical model of the supercell, we align 40 unit cells along the $y$-axis and assume the infinite periodic condition along the $x$-axis.
        \textbf{b} Dispersion diagrams of the QVH-based symmetry (black) and asymmetry (red) unit cell.
        The solid and open plots represent the out-of-plane and in-plane modes, respectively.
        \textbf{c} Side length $L_1$ of the triangular plate as a function of the $y$-axis coordinate for $\lambda=0.01$ \textmu m (circles), $\lambda=50$ \textmu m (crosses), and $\lambda=100$ \textmu m (triangles).
        Dispersion diagrams of the supercells having armchair boundary for (\textbf{d}) $\lambda=0.01$ \textmu m and (\textbf{e}) $\lambda=100$ \textmu m.
        The bulk band gap (shaded gray) and inner band gap (shaded green) depend on the asymmetric and adiabatic factors, respectively.
        \textbf{f} Bandwidth of the inner band gap $\Delta f$ as a function of adiabatic factor $\lambda$.
        The circle plot and dashed line represent the numerical and theoretical results, respectively.
    }
    \label{fig:figure1}
\end{figure}

We design our original unit cell consisting of two asymmetric triangular sub-lattices to investigate the topological nature in the QVH-based domain wall (Figure~\ref{fig:figure1}a).
The asymmetry between the two sub-lattices is introduced by the side lengths of the triangular plates $L_1$ and $L_2$.
The width of beams connecting the sub-lattices is $w=3$ \textmu m and the distance between centers of the nearest neighboring triangles is $a=60$ \textmu m.
The thickness of the whole structure is 1 \textmu m.
Assuming the infinite periodic condition of the unit cell, the dispersion diagrams for the out-of-plane modes (solid symbols) have a completely closing and opening bulk band gap at K point for $(L_1,L_2)=(32.5, 32.5)$ \textmu m (black line) and $(L_1,L_2)=(50, 10)$ \textmu m (red line), respectively, as shown in Figure~\ref{fig:figure1}b.
Note that the open symbols in Figure~\ref{fig:figure1}b indicate in-plane modes, which can be neglected when exciting the structure in the out-of-plane direction. 

We prepare the domain wall along the $x$-axis as shown in Figure~\ref{fig:figure1}a to fabricate the armchair boundary.
In our structure, the boundary is introduced by the inversion of the mass term corresponding to the side length of triangular plate:
\begin{align}
    \label{eq:1}
    L_{1/2}=L_0 \left[ 1\pm c \tanh \left( \frac{y_n}{\lambda} \right) \right],
\end{align}
where, $L_0(=32.5$ \textmu m$)$ is the side length for symmetric unit cell, $c(=0.538)$ is an asymmetric factor between the two sub-lattices, $y_n$ is the position of the $n_{\mathrm{th}}$ unit cell along the $y$-axis, and $\lambda$ is an adiabatic factor (mass domain wall width).
We design the supercell with not only $\lambda=0.01$ \textmu m as a step-like domain wall but also $10\leq \lambda \leq 300$ \textmu m to investigate the effect of adiabaticity on the armchair boundary.
Figure~\ref{fig:figure1}c shows $L_1$ corresponding to the mass term as function of $y_n$ for $\lambda=0.01$ (black circles), $50$ (red crosses), and $100$ (blue triangles) \textmu m.
We see that the profiles of $L_1$ around the domain wall change from a step-like to a smooth slope with increasing $\lambda$.

We first compare the dispersion diagrams of the supercells having different $\lambda$.
Figures~\ref{fig:figure1}d and \ref{fig:figure1}e show the dispersion diagrams of the supercells with the armchair boundary for $\lambda=0.01$ and $\lambda=100$ \textmu m, respectively.
In both dispersion diagrams, the red and blue lines indicate the interface modes of forward- and backward-propagation near $k_x=0$, respectively.
For a non-adiabatic boundary such as $\lambda=0.01$ \textmu m, inter-valley mixing between the K and K${}^\prime$ points opens a band gap in the interface mode dispersion within the bulk band gap (green shaded region). 
We denote the width of this interface mode minigap (inner gap) by $\Delta f$.
Such a gap opening is well known and reflects the valley-mixing mechanism responsible for the weak topological protection of QVH-based armchair boundaries.
For $\lambda=100$ \textmu m, the smooth domain wall suppresses the large Fourier components needed to scatter between K and K$^\prime$, thereby reducing inter-valley coupling. 
Consequently, the avoided crossing near $k_x \approx 0$ weakens and the inner band gap $\Delta f$ decreases.
Such inner band gap closing denotes the untanglement of the valley-mixing due to the adiabatic geometry, resulting in the enhancement of topological protection for the armchair boundary.

We theoretically investigate the improvement of the topological nature by the adiabaticity.
The bulk effective Hamiltonian around the K$^{(\prime)}$ point, far away from any interface, is given by
\begin{align}
    {\mathcal H}_{\tau_z}=
    -i v(\tau_z \sigma_1 \partial_x + \sigma_2 \partial_y) + \Delta G\sigma_3
    \label{eq:cont_hamiltonian}
    ,
\end{align}
where $v$ is the effective velocity,
$\partial_x$ and $\partial_y$ are the spatial derivative in the $x$ and $y$ direction, $\bm\sigma$ is the Pauli matrix vector, $\tau_z$ is the valley index, and $\Delta G$ is the band gap.
Next, we model the domain wall as
        \begin{align}
            \Delta G= \frac{\alpha v c}{\sqrt{3}a} \tanh({\bm e}\cdot {\bm r}/\lambda)
            ,
        \end{align}
        where
        $\alpha$ is a fitting parameter,
        $c$ is the asymmetric factor in Eq.~\eqref{eq:1},
        $\bm e$ is the direction normal to the topological boundary, and ${\bm r} = (x,y)$ is the lattice point coordinate.
        Then, we calculate the inner band gap arising at the armchair interface mode.
        This gap is given by \cite{Shah2021}
        \begin{align}
            \Delta f & = \sqrt{8 v^2 c /(\sqrt{3}\pi\lambda a)}\,  e^{-\mathcal{R} \lambda}
            \label{eq:Deltaf_theory}
            ,
        \end{align}
        where
        $\mathcal{R} = 2\int_{-\frac{2\pi}{9 a}}^{0} \arctan\left[\frac{1}{3c} \left(2 \cos\left(\frac{{3} \, k_x a}{2}\right)-1\right)\right]\, dk_x$.
        As the band gap of the armchair interface mode narrows, we expect that the topological protection of the interface mode recovers.

        The dashed line in Figure~\ref{fig:figure1}f plots Eq.~(\ref{eq:Deltaf_theory}), where $v$ is obtained from the dispersion relation in Figure~\ref{fig:figure1}e with fitting parameter $\alpha\approx 2$ at $\lambda = 100$ \textmu m.
$c=0.538$ is set by the numerical model,
and $k_x$ is the wavenumber parallel to the armchair edge.
We see that the energy gap in the band for the armchair boundary narrows as the adiabatic factor $\lambda$ (mass domain wall width) is increased.
Further, the theoretical result is in good agreement with the numerical calculation (red circle plots), ensuring that our armchair-based model is within the generalized theoretical framework of Dirac dispersion-based QVH.
The discrepancy between the theoretical and numerical plots for $\lambda<25$ \textmu m likely stems from the discreteness of the domain wall, which is not accounted for in the analytical model given by Eq.~(\ref{eq:cont_hamiltonian}). 
We also investigate the finite-size effect in our numerical model.
The localized profile of the topological interface modes generally broadens as $c$ decreases.
Thus, small $c$ leads to the finite-size-effect-induced discrepancy of $\Delta f$ in dispersion relations between simulation and theory.
Our results show that the finite-size effect is negligible for numerical models with $c>0.4$ and $\lambda=100$ \textmu m (section 1 of the supporting information).
Therefore, we select the validated adiabatic and asymmetric parameters of $\lambda=100$ \textmu m and $c=0.538$ as a topologically protected adiabatic armchair boundary for further numerical and experimental demonstrations.

\subsection{Topological protection in adiabatic armchair boundary}
In this section, we use a finite-structure transport simulations to show that a sufficiently adiabatic armchair domain wall suppresses intervalley scattering and recovers valley-protected (defect-tolerant) interface-mode propagation across most of the bulk gap. 

\begin{figure}[t!]
    \centering
    \includegraphics[width=1.0\textwidth]{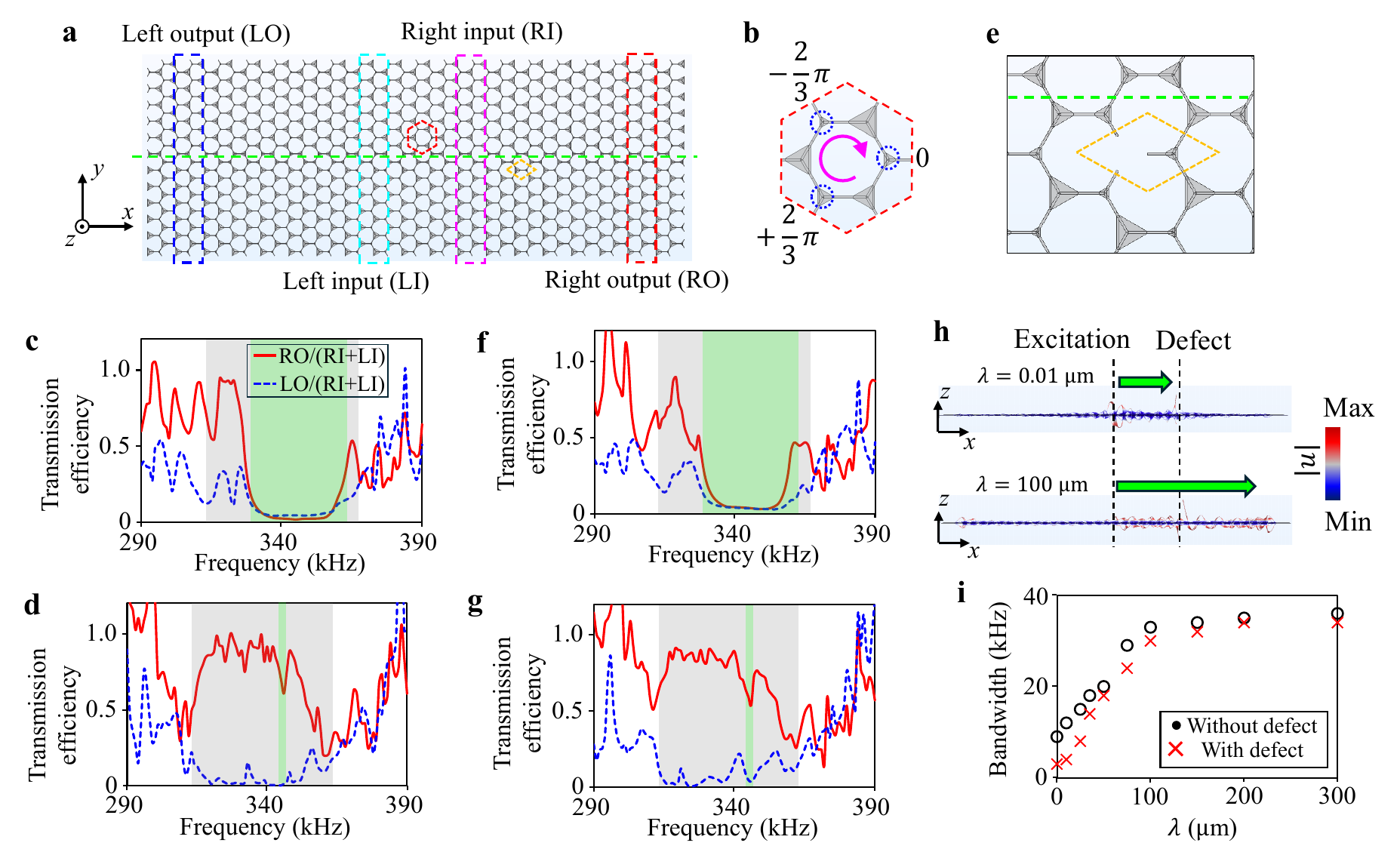}
    \caption{
        \textbf{Transmission efficiency and immunity to a defect of armchair boundary.}
        \textbf{a} The top view of a numerical model of a straight armchair waveguide (broken green line).
        The red hexagon and yellow rhombus represent the excitation point and position of the defect, respectively.
        In the waveguide, the input energy is calculated by summing kinetic energy at the right input (RI, magenta) and left input (LI, cyan) ports.
        The output energy of the rightward and leftward elastic waves is calculated at the right output (RO, red) and left output (LO, blue) ports, respectively.
        \textbf{b} The enlarged view of the excitation point.
        In this schematic, the three small sub-lattices are excited with a phase difference of $2\pi/3$ in a clockwise direction to excite the rightward wave.
        The transmission efficiencies of the straight armchair waveguides without the defect for (\textbf{c}) $\lambda=0.01$ and (\textbf{d}) $\lambda=100$ \textmu m.
        The solid red (dashed blue) line indicates the transmission efficiency from the input ports to the right (left) output port.
        The shaded gray and green regions represent the bulk and inner band gaps obtained by the dispersion diagrams of the unit cell.
        \textbf{e} The enlarged view of the sub-lattice defect in (a).
        The transmission efficiencies of the straight armchair waveguides with the defect for (\textbf{f}) $\lambda=0.01$ and (\textbf{g}) $\lambda=100$ \textmu m.
        \textbf{h} The displacement profiles on the $xz$-plane of the waveguides with the defect at 334 kHz for $\lambda=0.01$ and $\lambda=100$ \textmu m.
        The black dashed lines at the center and right of the waveguide represent the excitation point and defective point, respectively.
        \textbf{i} The bandwidth exceeding 70\% transmission efficiency as function of $\lambda$.
        The open circles and crosses show the results without and with the defect, respectively.
    }
    \label{fig:figure2}
\end{figure}

We verify the $\lambda$-dependent topological protection of the interface modes for the armchair boundary.
Figure~\ref{fig:figure2}a shows a numerical model of the straight armchair waveguide (broken green line).
We excite a region highlighted in a red hexagon by loading along the $z$-axis with sweeping frequency from 290 to 390 kHz.
The excitation point is enlarged in Figure~\ref{fig:figure2}b.
The three small sub-lattices in the hexagon are excited with a phase difference of $2\pi/3$ in a clockwise direction.
We calculate the total kinetic energy at the right input (RI: magenta rectangle), right output (RO: red rectangle), left input (LI: cyan rectangle), and left output (LO: blue rectangle) ports.
In this waveguide, we define the kinetic energy ratio RO/(RI+LI) [LO/(RI+LI)] as rightward [leftward] propagation efficiency.

Figure~\ref{fig:figure2}c and \ref{fig:figure2}d show the frequency spectra of rightward (solid red line) and leftward (dashed blue line) propagation efficiencies of the waveguides for $\lambda=0.01$ \textmu m and $100$ \textmu m, respectively.
In the spectra, the shaded gray and green regions express the bulk band gap and inner band gap based on the dispersion diagrams in Figures~\ref{fig:figure1}d and \ref{fig:figure1}e.
For $\lambda=0.01$ \textmu m, we see the highly efficient rightward propagation at the low frequency within the bulk band gap, while the efficiency drastically decreases in the inner band gap and the high frequency region of the bulk band gap.
Indeed, our numerical phase maps of the supercell having an armchair boundary with $\lambda=0.01$ clearly show that the interface mode at low (high) frequency region within the bulk band gap exhibits well (ill) propagating wave (section 2 of the supporting information).
On the other hand, for $\lambda=100$ \textmu m, the rightward interface modes maintain high efficiency across a wide range in the bulk band gap.
In addition, we see that the high-frequency interface mode within the bulk band gap also exhibits a propagating wave as topological protection becomes stronger with increasing $\lambda$ (section 2 of the supporting information).
Note that, for both cases of $\lambda$, the output at port LO remains at a low level in the entire range of the bulk band gap.

To further show the existence of topological protection in the adiabatic armchair waveguide, we investigate $\lambda$-dependent immunity to a defect by introducing a vacancy, as highlighted by the yellow rhombus in Figures~\ref{fig:figure2}a and \ref{fig:figure2}e.
Figures~\ref{fig:figure2}f and \ref{fig:figure2}g show the frequency spectra of transmission efficiency of the waveguides having the defect for $\lambda=0.01$ \textmu m and $100$ \textmu m, respectively.
In comparison with Figure~\ref{fig:figure2}c, Figure~\ref{fig:figure2}f indicates that the rightward transmission efficiency deteriorates at low frequencies within the bulk band gap.
Accordingly, the non-adiabatic armchair boundary is vulnerable to structural defects.
On the other hand, for $\lambda=100$ \textmu m, the transmission efficiency remains high and is essentially unchanged by the defect, as shown in Figure~\ref{fig:figure2}g.
The strong immunity to defect is one of evidences of the untanglement of valley-mixing by the adiabatic approach.
As one of the examples of wave propagation in the waveguides, Figure~\ref{fig:figure2}h visualizes displacement profiles along the waveguides at 334 kHz for $\lambda=0.01$ \textmu m and $100$ \textmu m, respectively.
We find that the defect suppresses elastic wave propagation in the waveguide for $\lambda=0.01$ \textmu m, whereas for $\lambda=100$ \textmu m the elastic wave passes through the defect and reaches the right end of the waveguide.

To understand the improvement mechanism of topological protection by the adiabatic geometry, we quantitatively evaluate the bandwidth for which the transmission efficiency exceeds 70\% within the bulk band gap.
Figure~\ref{fig:figure2}i shows the bandwidth as a function of $\lambda$ for the waveguide without (black circles) and with (red crosses) the defect.
We find that the bandwidth increases monotonically with increasing $\lambda$ regardless of the presence of the defect and is saturated for $\lambda\geq100$~\textmu m.
Upon careful examination, the increase in bandwidth appears to proceed in two distinct steps.
For $\lambda\leq50$~\textmu m, the transmission efficiency increases in the low frequency region within the bulk band gap with increasing $\lambda$, and
for $\lambda>50$~\textmu m, the transmission efficiency also begins to rise in the higher frequency region.
Then, the bandwidth almost saturates for $\lambda\geq100$~\textmu m because the interface modes exhibit high transfer efficiency across the bulk band gap.
Note that the saturated value of bandwidth corresponds to 80\% of the entire bandwidth of the bulk band gap.
The details of the adiabaticity-induced topological protection described above are discussed in Section 3 of the supporting information.
Comparing the bandwidth of the waveguides with and without the defect, the discrepancy between the two results decreases with increasing $\lambda$.
Therefore, in terms of the immunity to defects, our results confirm that an adiabatic geometry significantly enhances topological protection by suppressing valley-mixing.

\subsection{Topologically protected armchair waveguides}
\begin{figure}[b!]
    \centering
    \includegraphics[width=1.0\textwidth]{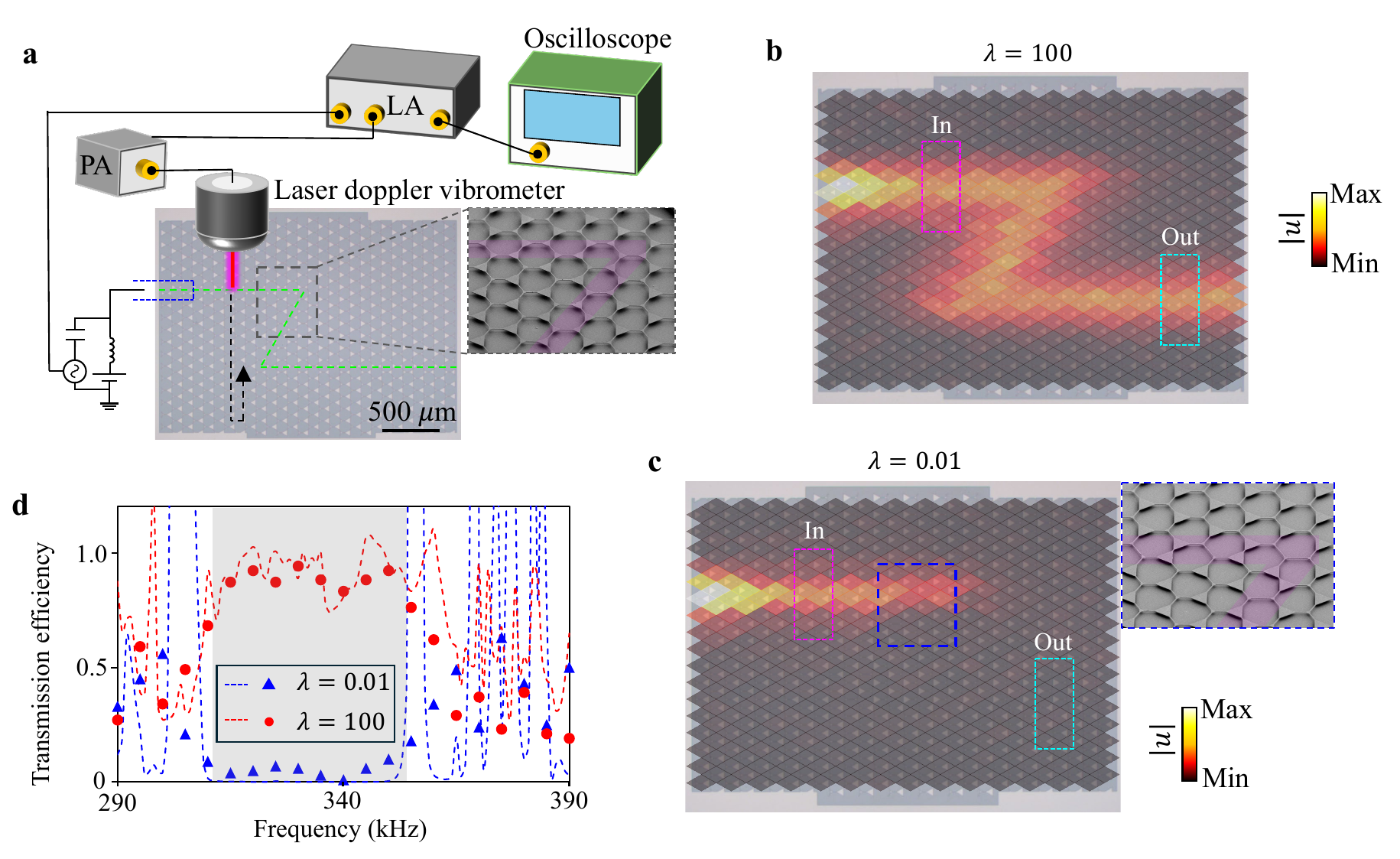}
    \caption{
        \textbf{Experimental setup and the wave propagation of armchair-based 120$^{\circ}$-bent waveguides.}
        \textbf{a} Top overview of the 120$^{\circ}$-bent armchair waveguide for $\lambda=100$ \textmu m and the schematic of the experimental setup to measure the vibration on the waveguide.
        The waveguide (dashed green line) is excited from the left edge of the waveguide by an excitation electrode on the back side of the substrate (dashed blue line).
        Here, a pre-amplifier and lock-in amplifier are described as PA and LA, respectively.
        The inset indicates the scanning electron microscope (SEM) image of the bending point in the waveguide (shaded magenta region).
        The 2D displacement profiles of the structures having the armchair waveguide for (\textbf{b}) $\lambda=100$ and (\textbf{c}) $\lambda=0.01$ \textmu m.
        The inset of (c) represents the SEM image around the bending point in the waveguide (shaded magenta line).
        \textbf{d} Transmission efficiencies of the armchair waveguides for $\lambda=100$ (red) and $\lambda=0.01$ (blue) \textmu m.
        The experimental results (symbols) are supported well by numerical results (dashed lines).
        The shaded gray region represents the bulk band gap obtained by the dispersion diagram of the supercell.
    }
    \label{fig:figure3}
\end{figure}

Motivated by the discussion above, we experimentally demonstrate sharply bent armchair waveguides that exhibit robust topological protection across the full bandwidth of the bulk band gap.
For experimental measurements, we prepare Si-based topological elastic waveguides by the fabrication processes similar to those in our previous work\cite{Funayama2021} (Method section).
Figure~\ref{fig:figure3}a shows the measurement setup and overview of a fabricated 2D MEMS structure.
A broken green line in the photo image represents the adiabatic armchair boundary-based topological waveguide.
As shown in the scanning electron micrograph (SEM) image in Figure~\ref{fig:figure3}a, the 120$^\circ$-bent armchair waveguide (shaded magenta line) consists of unit cells that closely match the numerical model for $\lambda=100$ \textmu m.
We excite the interface modes at the left end of the waveguide by using an excitation electrode (highlighted by the blue line) on the back side of the waveguide.
The vibration of each unit cell in the 2D-structure is measured via a scanning laser Doppler vibrometer.
The output signal from the laser Doppler vibrometer is amplified and filtered by a pre-amplifier (PA) and lock-in amplifier (LA), and then measured by an oscilloscope.

Figure~\ref{fig:figure3}b shows the 2D profile of out-of-plane displacement $u$ on the armchair boundary-based waveguide for $\lambda=100$ \textmu m at 330 kHz.
We find that the excited elastic wave propagates from left to right on the waveguide through the two bending points.
Such wave propagation is a unique characteristic of topological waveguides with spin-momentum locking, which suppresses reflection and backscattering.
On the other hand, the elastic wave can not pass through the first bent corner in the waveguide for $\lambda=0.01$ \textmu m at 330 kHz as shown in Figure~\ref{fig:figure3}c because of the non-negligible inter-valley scattering.
We calculate the values of the sum of $u^2$ in the magenta and cyan rectangles as input and output energy of the waveguides in Figures~\ref{fig:figure3}b and \ref{fig:figure3}c, respectively, and evaluate the transmission efficiencies by their ratio.

Figure~\ref{fig:figure3}d shows the experimental frequency spectra of transmission efficiency of the 120$^{\circ}$-bent waveguides for $\lambda=100$ (red circles) and $0.01$ (blue triangles) \textmu m, with each of the numerical results by the same colored lines as the experimental plots.
For $\lambda=100$ \textmu m, the transmission efficiency remains high, exceeding 80\% over most of the bulk band gap (shaded gray region).
As a result, the 120$^{\circ}$-bent adiabatic armchair waveguide exhibits performance comparable to the conventionally reported QVH-waveguides composed of the boundaries along the $\Gamma$K direction, i.e., zigzag and bridge boundaries\cite{WuX2017,Tian2020,LuJ2017,ChenL2022,HuoS2017}.
In contrast, the waveguide for $\lambda=0.01$ \textmu m shows low efficiency in the bulk band gap.
This result indicates that the non-adiabatic armchair boundary induces substantial backscattering due to the mixing of the wavefunction at the K and K$^{\prime}$ points.

The QVH-based armchair boundary provides us with further developments for the designability and functionality of topological waveguides.
By combining the armchair boundary with the other boundaries, i.e., the zigzag and bridge boundaries, QVH-based waveguides other than 120$^\circ$-bent waveguides are formable.
So far, such a scheme has been demonstrated toward universally designable topological waveguides\cite{WongS2020}.
However, the non-adiabatic armchair boundary has prevented the waves from propagating through the connection point between different boundaries.

In addition to the recovery of topological protection by tuning the adiabaticity, the matching of wavenumbers and frequencies between two waveguide modes is another significant factor for reducing the backscattering at the connection point.
When the interface modes of the adiabatic armchair and zigzag (bridge) boundaries satisfy the phase matching, energy and momentum conservations hold simultaneously by the periodicity of each boundary geometry, i.e., $\omega_{\mathrm{zig(bri)}}=\omega_{\mathrm{arm}}$ and $k_{\mathrm{zig(bri)}}+2\pi m/P_{\mathrm{zig(bri)}}=k_{\mathrm{arm}}+2\pi n/P_{\mathrm{arm}}$, where $\omega_{\mathrm{B}}$ and $k_{\mathrm{B}}$ represent frequency and wavenumber, $P_{\mathrm{B}}$ is the periodicity of boundary geometry, subscript indicates kinds of boundaries, $m$ and $n$ represent integer numbers.
For the connection between the armchair and zigzag (bridge) boundaries, the case where $m=1$ and $n=0$ is one of the combinations of $m$ and $n$ satisfying the momentum conservation:
$k_{\mathrm{zig}}=k_{\mathrm{arm}}-2\pi/P_{\mathrm{zig}}$, where the periodicity of zigzag boundary is $P_{\mathrm{zig}}=\sqrt{3}a$ in our devices.
Indeed, when we plot the dispersion diagrams for the zigzag (bridge) and the adiabatic armchair boundaries while considering the periodicity, we see that the interface modes in both dispersion diagrams match well across the bulk band gap (section 4 of the supporting information).
On the other hand, the interface modes for the non-adiabatic armchair have a distorted dispersion relation, unlike Dirac dispersion, resulting in a mismatch with the interface modes in the dispersion relation for zigzag (bridge) boundaries, regardless of the periodic term.
Accordingly, both effects of the adiabaticity and periodicity are crucial to enable phase matching of interface modes toward coupling between the armchair and the other boundaries.
Thus, our adiabatic armchair boundary has the potential to pave the way for universally designable QVH-based topological waveguides exhibiting topological protection within the bulk band gap.
We prepare the adiabatic armchair boundary-based waveguides combined with the zigzag and bridge boundaries to experimentally demonstrate 90$^{\circ}$- and 150$^{\circ}$-bent topological waveguides, respectively.

\begin{figure}[b!]
    \centering
    \includegraphics[width=1.0\textwidth]{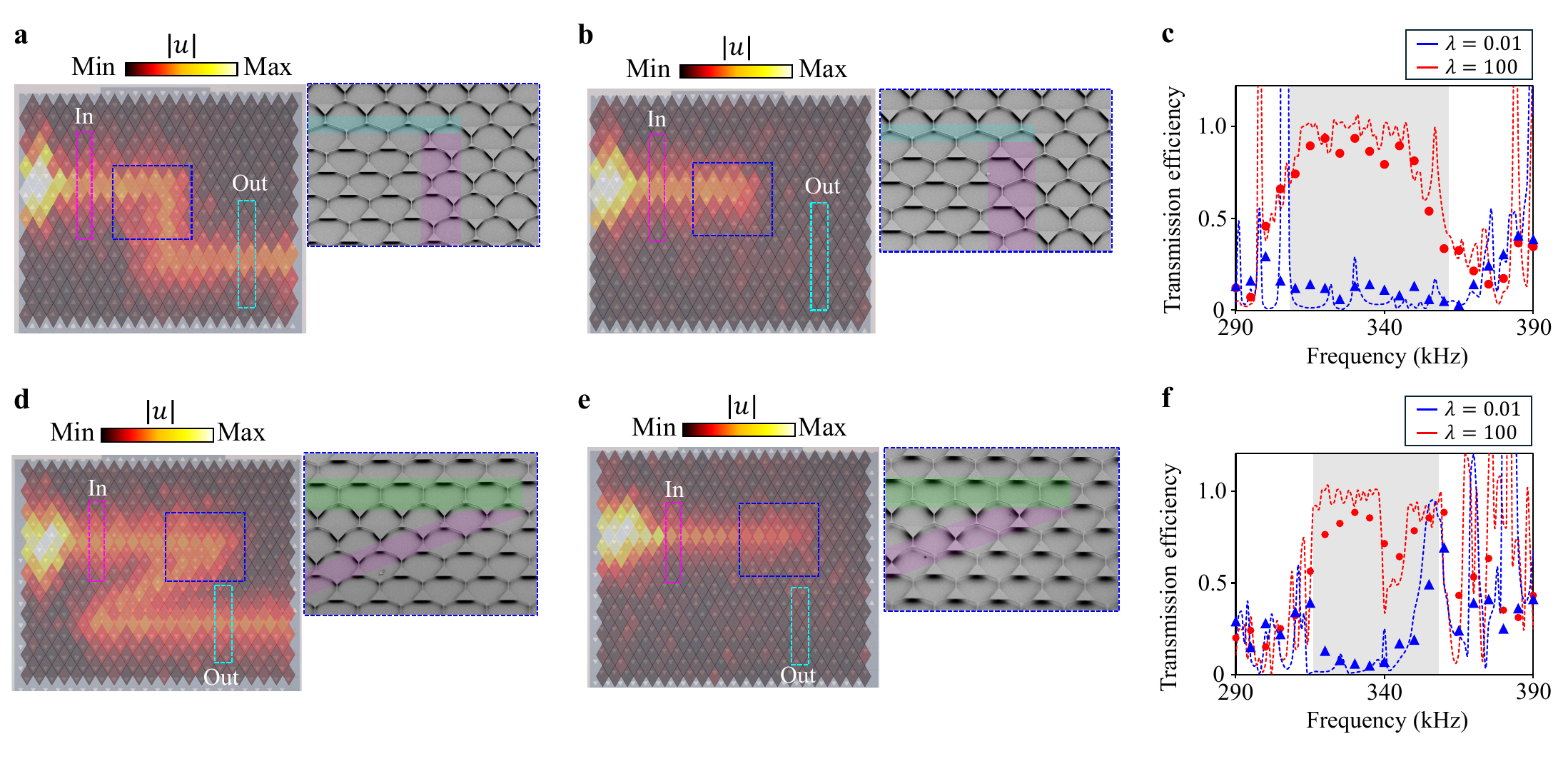}
    \caption{
        \textbf{Wave propagation through armchair boundary combined with other boundaries.}
        The 2D displacement profiles of the zigzag waveguide combined with the armchair waveguide for (\textbf{a}) $\lambda=100$ and (\textbf{b}) $\lambda=0.01$ \textmu m.
        Each inset in both figures shows the scanning electron microscopic (SEM) image around the first bending point highlighted by blue squares.
        The shaded blue and magenta regions in the SEM images represent the zigzag and armchair boundaries, respectively.
        \textbf{c} Transmission efficiencies of the waveguides including the armchair boundary for $\lambda=100$ (red) and $\lambda=0.01$ (blue) \textmu m.
        Those experimental transmission efficiencies are calculated by the ratio of squared displacement at the input port (magenta rectangle in (a) and (b)) to the output port (cyan rectangle in (a) and (b)).
        The symbols and lines indicate the experimental and numerical results, respectively.
        The shaded gray region shows the common bulk band gap of the zigzag and armchair boundaries.
        The 2D displacement profiles of the bridge waveguide combined with the armchair waveguide for (\textbf{d}) $\lambda=100$ and (\textbf{e}) $\lambda=0.01$.
        In the inset of the SEM images, the shaded green region represents the bridge boundary.
        \textbf{f} Transmission efficiencies of the waveguides in (d) and (e) calculated similarly to the results in (c).
    }
    \label{fig:figure4}
\end{figure}

First, Figures~\ref{fig:figure4}a and \ref{fig:figure4}b show the 2D displacement profiles at 330 kHz and SEM images of 90$^{\circ}$-bent waveguides for $\lambda=100$ \textmu m and $\lambda=0.01$ \textmu m, respectively.
The bending points are the connections between the zigzag (shaded cyan region) and armchair (shaded magenta region) boundaries, as shown in the SEM images.
In those waveguides, the elastic waves are excited at the left end of the zigzag waveguides, similar to the excitation of the 120$^{\circ}$-bent armchair waveguides.
According to the 2D displacement profile for $\lambda=100$ \textmu m, we see that the elastic wave propagates through the waveguide via the connecting points satisfying phase matching between the zigzag and armchair boundaries and topological protection.

Exciting the waveguide for $\lambda=0.01$ \textmu m at the same frequency as that for $\lambda=100$ \textmu m, the elastic wave propagates along the first zigzag boundary, whereas the wave amplitude rapidly decays near the first connecting point.
This wave attenuation is due to the phase mismatch at the connection between the two waveguides argued above, as well as wave reflection and scattering at the bend.
Consequently, little to no propagation is observed in the armchair section or in the second zigzag waveguides.
Figure~\ref{fig:figure4}c shows a frequency spectrum of transmission efficiency of the 90$^{\circ}$-bent waveguides.
The waveguide, including the adiabatic armchair boundary, achieves over 80\% efficiency across a wide range of the bulk band gap (shaded gray region).
On the other hand, low efficiency is observed in the waveguide with zigzag and non-adiabatic armchair boundaries.

As the second demonstration, we prepare waveguides having extremely sharp corners, i.e., 150$^{\circ}$-bent waveguides, by combining the bridge and armchair boundaries.
Figures~\ref{fig:figure4}d and \ref{fig:figure4}e show the displacement profiles and SEM images of the waveguides for $\lambda=100$ \textmu m and $\lambda=0.01$ \textmu m, respectively.
The SEM images enlarge the first bending point, showing the bridge (shaded green) and armchair (shaded magenta) boundaries.
In the waveguide with the adiabatic armchair boundary ($\lambda=100$ \textmu m), we find that the excited elastic wave propagates from left to right without significant loss.
Thus, the bridge and armchair boundaries couple each other due to phase matching, and suppress wave reflection and scattering by topological protection similar to the waveguide consisting of the armchair and zigzag boundaries.
On the other hand, when the waveguide includes the non-adiabatic armchair boundary ($\lambda=0.01$ \textmu m), the elastic wave propagating along the first bridge boundary almost dissipates at the connecting point with the armchair boundary by phase mismatching and weak topological protection.
We evaluate the frequency spectra of transmission efficiency in the waveguides as shown in Figure~\ref{fig:figure4}f.
The 150$^{\circ}$-bent waveguide, including the armchair boundary for $\lambda=100$ \textmu m, exhibits a higher efficiency than 80\% across the bulk band gap (shaded gray region).
Note that the efficiency reduction at the band edge is caused by weak topological protection.
When the 150$^{\circ}$-bent waveguide has the armchair boundary for $\lambda=0.01$ \textmu m, the wave propagation is suppressed, resulting in low transmission efficiency within the bulk band gap.
Therefore, our demonstrations clearly show that the adiabatic armchair boundary enables the untanglement of valley-mixing and the QVH-based universal topological waveguides.
Note that the increase in the transmission efficiency at the high-frequency region within the bulk band gap even for $\lambda=0.01$ \textmu m is due to the direct coupling between the two parallel bridge boundaries of IN and OUT.
In the bridge boundary, the interface mode profiles have been reported to broaden with increasing frequency\cite{Funayama202412,Funayama2025}, resulting in the direct tunneling between the adjacent two parallel bridge boundaries from IN and OUT without propagation through the armchair boundary (section 5 of the supporting information).

\section{Conclusion}
We have investigated adiabatic armchair waveguides to untangle valley mixing and enable universally designable QVH-based waveguides.
The effect of adiabatic geometry on valley mixing was investigated theoretically and numerically, and the findings were experimentally verified.
Our studies show that the width of the valley-mixing-induced inner band bap $\Delta f$ within the bulk band gap decreases with increasing adiabatic factor $\lambda$.
By untangling valley-mixing, the band gap for the interface modes is reduced, resulting in an improvement of topological protection for wave propagation.
Based on these theoretical and numerical results, we found that the adiabatic geometry improved the transmission efficiency of the adiabatic waveguide across the entire bulk band gap due to topological protection.
Our MEMS-based structure experimentally demonstrated that the 120$^{\circ}$-bent adiabatic armchair waveguide achieved as high transmission efficiency as the conventionally reported zigzag and bridge-based topological waveguides.
Furthermore, we experimentally demonstrated QVH-based waveguides with an adiabatic armchair boundary and other boundaries.
These different boundaries can be connected owing to the adiabaticity-induced phase matching and enhanced topological protection, enabling the 90$^{\circ}$- and 150$^{\circ}$-bent topological waveguides.
Overall, our results open the way to universal topological designs and expand the potential of topological devices for controlling classical and quantum wave phenomena and diffusion.

\section{Methods}
\subsection{Numerical simulation}
Our numerical models were built in COMSOL Multiphysics 6.2 with MEMS module.
For the calculations of dispersion diagrams of the unit cell, we set the Floquet-periodic boundary condition at all cross sections of the beams connecting the nearest neighbor sub-lattices.
For the numerical supercell, 40 unit cells were aligned perpendicular to the armchair boundary.
We set a finite periodic condition along the direction horizontal to the boundary.
Side lengths of the sub-lattices $L_\mathrm{1}$ and $L_\mathrm{2}$ were calculated as Eq.~\eqref{eq:1} with fixing $L_{0}$ and $c$ to $32.5$ \textmu m and $0.538$.
The dispersion diagrams of the models above were calculated by eigenvalue analysis in the module with sweeping wave vector.

The numerical models of straight armchair waveguides consisted of 38$\times$12 unit cells.
The positions of the small and large sub-lattices are inverted between the upper 6 and lower 6 units cells.
All edges of the structure were set to a low reflection boundary condition.
We swept the excitation frequency from 290 kHz to 390 kHz in 1 kHz steps.

When we simulated the 120$^{\circ}$-, 90$^{\circ}$-, and 150$^{\circ}$-bent waveguides, we built the structure completely mimicking the experimental structures.
The upper-left, upper-right, lower-left, and lower-right of the structures were set to fixed end conditions to mimic the four corners of the experimental structure connected to the Si device layer to suspend the waveguide.
We excited the 4 unit cells at the left end of the waveguides without the phase difference, such as the case of Figure~\ref{fig:figure2}b, to replicate the experimental excitation.

\subsection{Device fabrication and experimental measurements}
We used a single-crystal silicon-on-insulator (SOI) substrate to fabricate the MEMS-based topological elastic waveguides.
The SOI substrate consists of 1 \textmu m silicon device layer, 10 \textmu m silicon dioxide layer, and 470 \textmu m silicon support layer.
At the first step, we described the pattern of the periodic structure having topological boundaries on the silicon device layer using electron beam (EB) lithography (JEOL JBX-6300FS) with a resist (ZEP520A, Zeon Corporation).
After pattern developing, the silicon device layer was etched by the Bosh process (MUC-21 ASE-Pegasus, Sumitomo Precision Products) to form the 2D structures.
At the next step, we patterned the excitation electrode on the silicon support layer using photolithography (EVG620, EVG) with photo resist (THMR-iP3100MM, Tokyo Ohka Kogyo Co., Ltd.).
Then, we developed the pattern and etched the silicon support layer around the excitation electrode to isolate them.
We released the 2D structure from the substrate layer by etching the silicon dioxide layer in hydrofluoric acid.
After supercritical drying the devices, we obtained the completed devices.

We measured the fabricated devices in a high-vacuum condition ($\leq2.0\times10^{-5}$ Pa) to suppress the damping effect of the atmosphere.
One manual probe was contacted with the silicon device layer.
We contacted another manual probe to the excitation electrode from the top contact hole formed by etching the silicon device layer and the silicon dioxide layer.
Thus, we could apply the electrostatic force between the silicon device layer and the support layer.
As the excitation signal, we applied a sinusoidal signal with a function generator (RIGOL, DG972) with offset bias via stabilized power supply (Kikusui Electronics Corporation, PMX110-0.6A) and bias tee (Tektronix Keithley Instruments, PSPL5530B).
The signal amplitude was 5.0 V, and the offset voltage was 40 V.
The out-of-plane vibration displacement was measured by a laser Doppler vibrometer (Ono Sokki, LV-1800).
For high-sensitivity measurement, the output signal from the vibrometer was amplified and filtered by a pre-amplifier (NF Corporation, CA5360) and a lock-in amplifier (NF Corporation, LI5660).
Then, a digital oscilloscope (Tektronix, MSO64B) displayed the displacement.
We controlled the measurement position of the 2D structures on the controllable four-axis piezo stages (Sigmakoki, VSGSP60(XY), VSGSP60(Z), and SHOT-304GS).

\section{Data availability}
All data for validating this paper are provided as a Source Data file. Additional data supporting this paper are available from the corresponding author upon request.

\section{Acknowledgements}
A part of this work was supported by Nagoya University microstructural characterization platform as a program of ‘Advanced Research Infrastructure for Materials and Nanotechnology in Japan (ARIM)’ of the Ministry of Education, Culture, Sports, Science and Technology (MEXT), Japan.

\section{Author contributions}
K.F. performed numerical simulations, fabrications, experiments, and analysis. J.N. performed a theoretical analysis. A.Y. supported the reinforcement of theoretical modeling and comprehensive understanding. All authors contributed to discussions and manuscript preparation.

\section{Competing interests}
The authors declare no competing interests.

\section{Additional information}
The Supporting Information is available free of charge.
Details of the asymmetric factor-dependent inner band gap width for the adiabatic armchair boundary, comparison of phase profiles of the supercells with and without adiabaticity, frequency spectra of the transmission efficiency, comparison of the dispersion diagrams of the different boundaries, and numerical displacement profiles representing the coupling between the bridge boundaries.


\begin{thebibliography}{57}
\ifx \bisbn   \undefined \def \bisbn  #1{ISBN #1}\fi
\ifx \binits  \undefined \def \binits#1{#1}\fi
\ifx \bauthor  \undefined \def \bauthor#1{#1}\fi
\ifx \batitle  \undefined \def \batitle#1{#1}\fi
\ifx \bjtitle  \undefined \def \bjtitle#1{#1}\fi
\ifx \bvolume  \undefined \def \bvolume#1{\textbf{#1}}\fi
\ifx \byear  \undefined \def \byear#1{#1}\fi
\ifx \bissue  \undefined \def \bissue#1{#1}\fi
\ifx \bfpage  \undefined \def \bfpage#1{#1}\fi
\ifx \blpage  \undefined \def \blpage #1{#1}\fi
\ifx \burl  \undefined \def \burl#1{\textsf{#1}}\fi
\ifx \doiurl  \undefined \def \doiurl#1{\url{https://doi.org/#1}}\fi
\ifx \betal  \undefined \def \betal{\textit{et al.}}\fi
\ifx \binstitute  \undefined \def \binstitute#1{#1}\fi
\ifx \binstitutionaled  \undefined \def \binstitutionaled#1{#1}\fi
\ifx \bctitle  \undefined \def \bctitle#1{#1}\fi
\ifx \beditor  \undefined \def \beditor#1{#1}\fi
\ifx \bpublisher  \undefined \def \bpublisher#1{#1}\fi
\ifx \bbtitle  \undefined \def \bbtitle#1{#1}\fi
\ifx \bedition  \undefined \def \bedition#1{#1}\fi
\ifx \bseriesno  \undefined \def \bseriesno#1{#1}\fi
\ifx \blocation  \undefined \def \blocation#1{#1}\fi
\ifx \bsertitle  \undefined \def \bsertitle#1{#1}\fi
\ifx \bsnm \undefined \def \bsnm#1{#1}\fi
\ifx \bsuffix \undefined \def \bsuffix#1{#1}\fi
\ifx \bparticle \undefined \def \bparticle#1{#1}\fi
\ifx \barticle \undefined \def \barticle#1{#1}\fi
\bibcommenthead
\ifx \bconfdate \undefined \def \bconfdate #1{#1}\fi
\ifx \botherref \undefined \def \botherref #1{#1}\fi
\ifx \url \undefined \def \url#1{\textsf{#1}}\fi
\ifx \bchapter \undefined \def \bchapter#1{#1}\fi
\ifx \bbook \undefined \def \bbook#1{#1}\fi
\ifx \bcomment \undefined \def \bcomment#1{#1}\fi
\ifx \oauthor \undefined \def \oauthor#1{#1}\fi
\ifx \citeauthoryear \undefined \def \citeauthoryear#1{#1}\fi
\ifx \endbibitem  \undefined \def \endbibitem {}\fi
\ifx \bconflocation  \undefined \def \bconflocation#1{#1}\fi
\ifx \arxivurl  \undefined \def \arxivurl#1{\textsf{#1}}\fi
\csname PreBibitemsHook\endcsname

\bibitem[\protect\citeauthoryear{Yan et~al.}{2021}]{W-Yan2021}
\begin{barticle}
\bauthor{\bsnm{Yan}, \binits{W.}},
\bauthor{\bsnm{Song}, \binits{D.}},
\bauthor{\bsnm{Xia}, \binits{S.}},
\bauthor{\bsnm{Xie}, \binits{J.}},
\bauthor{\bsnm{Tang}, \binits{L.}},
\bauthor{\bsnm{Xu}, \binits{J.}},
\bauthor{\bsnm{Chen}, \binits{Z.}}:
\batitle{Realization of second-order photonic square-root topological insulators}.
\bjtitle{ACS Photonics}
\bvolume{8}(\bissue{11}),
\bfpage{3308}--\blpage{3314}
(\byear{2021})
\doiurl{10.1021/acsphotonics.1c01171}
\end{barticle}
\endbibitem

\bibitem[\protect\citeauthoryear{Wang et~al.}{2025}]{Y-Wang2025}
\begin{barticle}
\bauthor{\bsnm{Wang}, \binits{Y.}},
\bauthor{\bsnm{Jia}, \binits{D.}},
\bauthor{\bsnm{Lu}, \binits{Y.-J.}},
\bauthor{\bsnm{Gu}, \binits{S.}},
\bauthor{\bsnm{Ge}, \binits{Y.}},
\bauthor{\bsnm{Yuan}, \binits{S.-Q.}},
\bauthor{\bsnm{Sun}, \binits{H.-X.}},
\bauthor{\bsnm{Yang}, \binits{Y.}},
\bauthor{\bsnm{Zhang}, \binits{B.}}:
\batitle{Acoustic pseudospin transport in dual-layer topological insulators with multiple topological phases}.
\bjtitle{Advanced Functional Materials}
\bvolume{35}(\bissue{29}),
\bfpage{2422858}
(\byear{2025})
\doiurl{10.1002/adfm.202422858}
\end{barticle}
\endbibitem

\bibitem[\protect\citeauthoryear{Zheng et~al.}{2022}]{Zheng2022}
\begin{barticle}
\bauthor{\bsnm{Zheng}, \binits{Z.}},
\bauthor{\bsnm{Yin}, \binits{J.}},
\bauthor{\bsnm{Wen}, \binits{J.}},
\bauthor{\bsnm{Yu}, \binits{D.}}:
\batitle{Higher-order topological states in locally resonant elastic metamaterials}.
\bjtitle{Applied Physics Letters}
\bvolume{120}(\bissue{14}),
\bfpage{144101}
(\byear{2022})
\doiurl{10.1063/5.0074463}
\end{barticle}
\endbibitem

\bibitem[\protect\citeauthoryear{Miniaci and Pal}{2021}]{Miniaci2021}
\begin{barticle}
\bauthor{\bsnm{Miniaci}, \binits{M.}},
\bauthor{\bsnm{Pal}, \binits{R.K.}}:
\batitle{Design of topological elastic waveguides}.
\bjtitle{Journal of Applied Physics}
\bvolume{130}(\bissue{14}),
\bfpage{141101}
(\byear{2021})
\doiurl{10.1063/5.0057288}
\end{barticle}
\endbibitem

\bibitem[\protect\citeauthoryear{Lin et~al.}{2021}]{Lin2021}
\begin{barticle}
\bauthor{\bsnm{Lin}, \binits{S.}},
\bauthor{\bsnm{Zhang}, \binits{L.}},
\bauthor{\bsnm{Tian}, \binits{T.}},
\bauthor{\bsnm{Duan}, \binits{C.-K.}},
\bauthor{\bsnm{Du}, \binits{J.}}:
\batitle{Dynamic observation of topological soliton states in a programmable nanomechanical lattice}.
\bjtitle{Nano Letters}
\bvolume{21}(\bissue{2}),
\bfpage{1025}--\blpage{1031}
(\byear{2021})
\doiurl{10.1021/acs.nanolett.0c04121}
\end{barticle}
\endbibitem

\bibitem[\protect\citeauthoryear{Yu et~al.}{2018}]{YuS2018}
\begin{barticle}
\bauthor{\bsnm{Yu}, \binits{S.-Y.}},
\bauthor{\bsnm{He}, \binits{C.}},
\bauthor{\bsnm{Wang}, \binits{Z.}},
\bauthor{\bsnm{Liu}, \binits{F.-K.}},
\bauthor{\bsnm{Sun}, \binits{X.-C.}},
\bauthor{\bsnm{Li}, \binits{Z.}},
\bauthor{\bsnm{Lu}, \binits{H.-Z.}},
\bauthor{\bsnm{Lu}, \binits{M.-H.}},
\bauthor{\bsnm{Liu}, \binits{X.-P.}},
\bauthor{\bsnm{Chen}, \binits{Y.-F.}}:
\batitle{Elastic pseudospin transport for integratable topological phononic circuits}.
\bjtitle{Nature Communications}
\bvolume{9}(\bissue{1}),
\bfpage{3072}
(\byear{2018})
\doiurl{10.1038/s41467-018-05461-5}
\end{barticle}
\endbibitem

\bibitem[\protect\citeauthoryear{Li et~al.}{2018}]{Y-Li2018}
\begin{barticle}
\bauthor{\bsnm{Li}, \binits{Y.}},
\bauthor{\bsnm{Sun}, \binits{Y.}},
\bauthor{\bsnm{Zhu}, \binits{W.}},
\bauthor{\bsnm{Guo}, \binits{Z.}},
\bauthor{\bsnm{Jiang}, \binits{J.}},
\bauthor{\bsnm{Kariyado}, \binits{T.}},
\bauthor{\bsnm{Chen}, \binits{H.}},
\bauthor{\bsnm{Hu}, \binits{X.}}:
\batitle{Topological lc-circuits based on microstrips and observation of electromagnetic modes with orbital angular momentum}.
\bjtitle{Nature Communications}
\bvolume{9}(\bissue{1}),
\bfpage{4598}
(\byear{2018})
\doiurl{10.1038/s41467-018-07084-2}
\end{barticle}
\endbibitem

\bibitem[\protect\citeauthoryear{Wang et~al.}{2009}]{WangZ2009}
\begin{barticle}
\bauthor{\bsnm{Wang}, \binits{Z.}},
\bauthor{\bsnm{Chong}, \binits{Y.}},
\bauthor{\bsnm{Joannopoulos}, \binits{J.D.}},
\bauthor{\bsnm{Soljačić}, \binits{M.}}:
\batitle{Observation of unidirectional backscattering-immune topological electromagnetic states}.
\bjtitle{Nature}
\bvolume{461}(\bissue{7265}),
\bfpage{772}--\blpage{775}
(\byear{2009})
\doiurl{10.1038/nature08293}
\end{barticle}
\endbibitem

\bibitem[\protect\citeauthoryear{Cha et~al.}{2018}]{ChaJ2018}
\begin{barticle}
\bauthor{\bsnm{Cha}, \binits{J.}},
\bauthor{\bsnm{Kim}, \binits{K.W.}},
\bauthor{\bsnm{Daraio}, \binits{C.}}:
\batitle{Experimental realization of on-chip topological nanoelectromechanical metamaterials}.
\bjtitle{Nature}
\bvolume{564}(\bissue{7735}),
\bfpage{229}--\blpage{233}
(\byear{2018})
\doiurl{10.1038/s41586-018-0764-0}
\end{barticle}
\endbibitem

\bibitem[\protect\citeauthoryear{Zilberberg et~al.}{2018}]{Zilberberg2018}
\begin{barticle}
\bauthor{\bsnm{Zilberberg}, \binits{O.}},
\bauthor{\bsnm{Huang}, \binits{S.}},
\bauthor{\bsnm{Guglielmon}, \binits{J.}},
\bauthor{\bsnm{Wang}, \binits{M.}},
\bauthor{\bsnm{Chen}, \binits{K.P.}},
\bauthor{\bsnm{Kraus}, \binits{Y.E.}},
\bauthor{\bsnm{Rechtsman}, \binits{M.C.}}:
\batitle{Photonic topological boundary pumping as a probe of 4d quantum hall physics}.
\bjtitle{Nature}
\bvolume{556}(\bissue{7686}),
\bfpage{59}
(\byear{2018})
\doiurl{10.1038/nature25011}
\end{barticle}
\endbibitem

\bibitem[\protect\citeauthoryear{Serra-Garcia et~al.}{2019}]{Serra2019}
\begin{barticle}
\bauthor{\bsnm{Serra-Garcia}, \binits{M.}},
\bauthor{\bsnm{S\"usstrunk}, \binits{R.}},
\bauthor{\bsnm{Huber}, \binits{S.D.}}:
\batitle{Observation of quadrupole transitions and edge mode topology in an lc circuit network}.
\bjtitle{Physical Review B}
\bvolume{99},
\bfpage{020304}
(\byear{2019})
\doiurl{10.1103/PhysRevB.99.020304}
\end{barticle}
\endbibitem

\bibitem[\protect\citeauthoryear{Zhao et~al.}{2019}]{ZhaoH2019}
\begin{barticle}
\bauthor{\bsnm{Zhao}, \binits{H.}},
\bauthor{\bsnm{Qiao}, \binits{X.}},
\bauthor{\bsnm{Wu}, \binits{T.}},
\bauthor{\bsnm{Midya}, \binits{B.}},
\bauthor{\bsnm{Longhi}, \binits{S.}},
\bauthor{\bsnm{Feng}, \binits{L.}}:
\batitle{Non-hermitian topological light steering}.
\bjtitle{Science}
\bvolume{365}(\bissue{6458}),
\bfpage{1163}--\blpage{1166}
(\byear{2019})
\doiurl{10.1126/science.aay1064}
\end{barticle}
\endbibitem

\bibitem[\protect\citeauthoryear{Hashemi et~al.}{2025}]{Hashemi2025}
\begin{barticle}
\bauthor{\bsnm{Hashemi}, \binits{A.}},
\bauthor{\bsnm{Zakeri}, \binits{M.J.}},
\bauthor{\bsnm{Jung}, \binits{P.S.}},
\bauthor{\bsnm{Blanco-Redondo}, \binits{A.}}:
\batitle{Topological quantum photonics}.
\bjtitle{APL Photonics}
\bvolume{10}(\bissue{1}),
\bfpage{010903}
(\byear{2025})
\doiurl{10.1063/5.0239265}
\end{barticle}
\endbibitem

\bibitem[\protect\citeauthoryear{He et~al.}{2024}]{HeL2024}
\begin{barticle}
\bauthor{\bsnm{He}, \binits{L.}},
\bauthor{\bsnm{Liu}, \binits{D.}},
\bauthor{\bsnm{Zhang}, \binits{H.}},
\bauthor{\bsnm{Zhang}, \binits{F.}},
\bauthor{\bsnm{Zhang}, \binits{W.}},
\bauthor{\bsnm{Feng}, \binits{X.}},
\bauthor{\bsnm{Huang}, \binits{Y.}},
\bauthor{\bsnm{Cui}, \binits{K.}},
\bauthor{\bsnm{Liu}, \binits{F.}},
\bauthor{\bsnm{Zhang}, \binits{W.}},
\bauthor{\bsnm{Zhang}, \binits{X.}}:
\batitle{Topologically protected quantum logic gates with valley-hall photonic crystals}.
\bjtitle{Advanced Materials}
\bvolume{36}(\bissue{24}),
\bfpage{2311611}
(\byear{2024})
\doiurl{10.1002/adma.202311611}
\end{barticle}
\endbibitem

\bibitem[\protect\citeauthoryear{Wang et~al.}{2019}]{WangM2019}
\begin{barticle}
\bauthor{\bsnm{Wang}, \binits{M.}},
\bauthor{\bsnm{Doyle}, \binits{C.}},
\bauthor{\bsnm{Bell}, \binits{B.}},
\bauthor{\bsnm{Collins}, \binits{M.J.}},
\bauthor{\bsnm{Magi}, \binits{E.}},
\bauthor{\bsnm{Eggleton}, \binits{B.J.}},
\bauthor{\bsnm{Segev}, \binits{M.}},
\bauthor{\bsnm{Blanco-Redondo}, \binits{A.}}:
\batitle{Topologically protected entangled photonic states}.
\bjtitle{Nanophotonics}
\bvolume{8}(\bissue{8}),
\bfpage{1327}--\blpage{1335}
(\byear{2019})
\doiurl{10.1515/nanoph-2019-0058}
\end{barticle}
\endbibitem

\bibitem[\protect\citeauthoryear{Barik et~al.}{2018}]{Sabyasachi2018}
\begin{barticle}
\bauthor{\bsnm{Barik}, \binits{S.}},
\bauthor{\bsnm{Karasahin}, \binits{A.}},
\bauthor{\bsnm{Flower}, \binits{C.}},
\bauthor{\bsnm{Cai}, \binits{T.}},
\bauthor{\bsnm{Miyake}, \binits{H.}},
\bauthor{\bsnm{DeGottardi}, \binits{W.}},
\bauthor{\bsnm{Hafezi}, \binits{M.}},
\bauthor{\bsnm{Waks}, \binits{E.}}:
\batitle{A topological quantum optics interface}.
\bjtitle{Science}
\bvolume{359}(\bissue{6376}),
\bfpage{666}--\blpage{668}
(\byear{2018})
\doiurl{10.1126/science.aaq0327}
\end{barticle}
\endbibitem

\bibitem[\protect\citeauthoryear{Blanco-Redondo et~al.}{2018}]{Balco2018}
\begin{barticle}
\bauthor{\bsnm{Blanco-Redondo}, \binits{A.}},
\bauthor{\bsnm{Bell}, \binits{B.}},
\bauthor{\bsnm{Oren}, \binits{D.}},
\bauthor{\bsnm{Eggleton}, \binits{B.J.}},
\bauthor{\bsnm{Segev}, \binits{M.}}:
\batitle{Topological protection of biphoton states}.
\bjtitle{Science}
\bvolume{362}(\bissue{6414}),
\bfpage{568}--\blpage{571}
(\byear{2018})
\doiurl{10.1126/science.aau4296}
\end{barticle}
\endbibitem

\bibitem[\protect\citeauthoryear{Hu et~al.}{2022}]{HuH2022}
\begin{barticle}
\bauthor{\bsnm{Hu}, \binits{H.}},
\bauthor{\bsnm{Han}, \binits{S.}},
\bauthor{\bsnm{Yang}, \binits{Y.}},
\bauthor{\bsnm{Liu}, \binits{D.}},
\bauthor{\bsnm{Xue}, \binits{H.}},
\bauthor{\bsnm{Liu}, \binits{G.-G.}},
\bauthor{\bsnm{Cheng}, \binits{Z.}},
\bauthor{\bsnm{Wang}, \binits{Q.J.}},
\bauthor{\bsnm{Zhang}, \binits{S.}},
\bauthor{\bsnm{Zhang}, \binits{B.}},
\bauthor{\bsnm{Luo}, \binits{Y.}}:
\batitle{Observation of topological edge states in thermal diffusion}.
\bjtitle{Advanced Materials}
\bvolume{34}(\bissue{31}),
\bfpage{2202257}
(\byear{2022})
\doiurl{10.1002/adma.202202257}
\end{barticle}
\endbibitem

\bibitem[\protect\citeauthoryear{Qi et~al.}{2022}]{QiM2022}
\begin{barticle}
\bauthor{\bsnm{Qi}, \binits{M.}},
\bauthor{\bsnm{Wang}, \binits{D.}},
\bauthor{\bsnm{Cao}, \binits{P.-C.}},
\bauthor{\bsnm{Zhu}, \binits{X.-F.}},
\bauthor{\bsnm{Qiu}, \binits{C.-W.}},
\bauthor{\bsnm{Chen}, \binits{H.}},
\bauthor{\bsnm{Li}, \binits{Y.}}:
\batitle{Geometric phase and localized heat diffusion}.
\bjtitle{Advanced Materials}
\bvolume{34}(\bissue{32}),
\bfpage{2202241}
(\byear{2022})
\doiurl{10.1002/adma.202202241}
\end{barticle}
\endbibitem

\bibitem[\protect\citeauthoryear{Wu et~al.}{2023}]{WuH2023}
\begin{barticle}
\bauthor{\bsnm{Wu}, \binits{H.}},
\bauthor{\bsnm{Hu}, \binits{H.}},
\bauthor{\bsnm{Wang}, \binits{X.}},
\bauthor{\bsnm{Xu}, \binits{Z.}},
\bauthor{\bsnm{Zhang}, \binits{B.}},
\bauthor{\bsnm{Wang}, \binits{Q.J.}},
\bauthor{\bsnm{Zheng}, \binits{Y.}},
\bauthor{\bsnm{Zhang}, \binits{J.}},
\bauthor{\bsnm{Cui}, \binits{T.J.}},
\bauthor{\bsnm{Luo}, \binits{Y.}}:
\batitle{Higher-order topological states in thermal diffusion}.
\bjtitle{Advanced Materials}
\bvolume{35}(\bissue{14}),
\bfpage{2210825}
(\byear{2023})
\doiurl{10.1002/adma.202210825}
\end{barticle}
\endbibitem

\bibitem[\protect\citeauthoryear{Funayama et~al.}{2024}]{FunayamaAPL2024}
\begin{barticle}
\bauthor{\bsnm{Funayama}, \binits{K.}},
\bauthor{\bsnm{Hirotani}, \binits{J.}},
\bauthor{\bsnm{Tanaka}, \binits{H.}}:
\batitle{Evaluation of topological protection in kagome lattice-based thermal diffusion systems}.
\bjtitle{Applied Physics Letters}
\bvolume{124}(\bissue{23}),
\bfpage{232201}
(\byear{2024})
\doiurl{10.1063/5.0214412}
\end{barticle}
\endbibitem

\bibitem[\protect\citeauthoryear{Funayama et~al.}{2023}]{Funayama2023}
\begin{barticle}
\bauthor{\bsnm{Funayama}, \binits{K.}},
\bauthor{\bsnm{Hirotani}, \binits{J.}},
\bauthor{\bsnm{Miura}, \binits{A.}},
\bauthor{\bsnm{Tanaka}, \binits{H.}}:
\batitle{Selectable diffusion direction with topologically protected edge modess}.
\bjtitle{Communications Physics}
\bvolume{6}(\bissue{1}),
\bfpage{364}
(\byear{2023})
\doiurl{10.1038/s42005-023-01490-9}
\end{barticle}
\endbibitem

\bibitem[\protect\citeauthoryear{Liu et~al.}{2024}]{LiuZ2024}
\begin{barticle}
\bauthor{\bsnm{Liu}, \binits{Z.}},
\bauthor{\bsnm{Jin}, \binits{P.}},
\bauthor{\bsnm{Lei}, \binits{M.}},
\bauthor{\bsnm{Wang}, \binits{C.}},
\bauthor{\bsnm{Marchesoni}, \binits{F.}},
\bauthor{\bsnm{Jiang}, \binits{J.-H.}},
\bauthor{\bsnm{Huang}, \binits{J.}}:
\batitle{Topological thermal transport}.
\bjtitle{Nature Reviews Physics}
\bvolume{6}(\bissue{9}),
\bfpage{554}
(\byear{2024})
\doiurl{10.1038/s42254-024-00745-w}
\end{barticle}
\endbibitem

\bibitem[\protect\citeauthoryear{Fukui et~al.}{2023}]{Fukui2023}
\begin{barticle}
\bauthor{\bsnm{Fukui}, \binits{T.}},
\bauthor{\bsnm{Yoshida}, \binits{T.}},
\bauthor{\bsnm{Hatsugai}, \binits{Y.}}:
\batitle{Higher-order topological heat conduction on a lattice for detection of corner states}.
\bjtitle{Physical Review E}
\bvolume{108},
\bfpage{024112}
(\byear{2023})
\doiurl{10.1103/PhysRevE.108.024112}
\end{barticle}
\endbibitem

\bibitem[\protect\citeauthoryear{Yoshida and Hatsugai}{2021}]{Yoshida2021}
\begin{barticle}
\bauthor{\bsnm{Yoshida}, \binits{T.}},
\bauthor{\bsnm{Hatsugai}, \binits{Y.}}:
\batitle{Bulk-edge correspondence of classical diffusion phenomena}.
\bjtitle{Scientific Reports}
\bvolume{11},
\bfpage{888}
(\byear{2021})
\doiurl{10.1038/s41598-020-80180-w}
\end{barticle}
\endbibitem

\bibitem[\protect\citeauthoryear{Li et~al.}{2019}]{LiY2019}
\begin{barticle}
\bauthor{\bsnm{Li}, \binits{Y.}},
\bauthor{\bsnm{Peng}, \binits{Y.-G.}},
\bauthor{\bsnm{Han}, \binits{L.}},
\bauthor{\bsnm{Miri}, \binits{M.-A.}},
\bauthor{\bsnm{Li}, \binits{W.}},
\bauthor{\bsnm{Xiao}, \binits{M.}},
\bauthor{\bsnm{Zhu}, \binits{X.-F.}},
\bauthor{\bsnm{Zhao}, \binits{J.}},
\bauthor{\bsnm{Alù}, \binits{A.}},
\bauthor{\bsnm{Fan}, \binits{S.}},
\bauthor{\bsnm{Qiu}, \binits{C.-W.}}:
\batitle{Anti–parity-time symmetry in diffusive systems}.
\bjtitle{Science}
\bvolume{364}(\bissue{6436}),
\bfpage{170}--\blpage{173}
(\byear{2019})
\doiurl{10.1126/science.aaw6259}
\end{barticle}
\endbibitem

\bibitem[\protect\citeauthoryear{Dong et~al.}{2017}]{DongJ2017}
\begin{barticle}
\bauthor{\bsnm{Dong}, \binits{J.-W.}},
\bauthor{\bsnm{Chen}, \binits{X.-D.}},
\bauthor{\bsnm{Zhu}, \binits{H.}},
\bauthor{\bsnm{Wang}, \binits{Y.}},
\bauthor{\bsnm{Zhang}, \binits{X.}}:
\batitle{Valley photonic crystals for control of spin and topology}.
\bjtitle{Nature Materials}
\bvolume{16}(\bissue{3}),
\bfpage{298}--\blpage{302}
(\byear{2017})
\doiurl{10.1038/nmat4807}
\end{barticle}
\endbibitem

\bibitem[\protect\citeauthoryear{Lu et~al.}{2017}]{LuJ2017}
\begin{barticle}
\bauthor{\bsnm{Lu}, \binits{J.}},
\bauthor{\bsnm{Qiu}, \binits{C.}},
\bauthor{\bsnm{Ye}, \binits{L.}},
\bauthor{\bsnm{Fan}, \binits{X.}},
\bauthor{\bsnm{Ke}, \binits{M.}},
\bauthor{\bsnm{Zhang}, \binits{F.}},
\bauthor{\bsnm{Liu}, \binits{Z.}}:
\batitle{Observation of topological valley transport of sound in sonic crystals}.
\bjtitle{Nature Physics}
\bvolume{13}(\bissue{4}),
\bfpage{369}
(\byear{2017})
\doiurl{10.1038/nphys3999}
\end{barticle}
\endbibitem

\bibitem[\protect\citeauthoryear{Jia et~al.}{2025}]{Jia2025}
\begin{barticle}
\bauthor{\bsnm{Jia}, \binits{R.}},
\bauthor{\bsnm{Tan}, \binits{Y.J.}},
\bauthor{\bsnm{Navaratna}, \binits{N.}},
\bauthor{\bsnm{Kumar}, \binits{A.}},
\bauthor{\bsnm{Singh}, \binits{R.}}:
\batitle{Photonic supercoupling in silicon topological waveguides}.
\bjtitle{Advanced Materials}
\bvolume{37}(\bissue{6}),
\bfpage{2415083}
(\byear{2025})
\doiurl{10.1002/adma.202415083}
\end{barticle}
\endbibitem

\bibitem[\protect\citeauthoryear{Xue et~al.}{2021}]{XueH2021}
\begin{barticle}
\bauthor{\bsnm{Xue}, \binits{H.}},
\bauthor{\bsnm{Yang}, \binits{Y.}},
\bauthor{\bsnm{Zhang}, \binits{B.}}:
\batitle{Topological valley photonics: Physics and device applications}.
\bjtitle{Advanced Photonics Research}
\bvolume{2}(\bissue{8}),
\bfpage{2100013}
(\byear{2021})
\doiurl{10.1002/adpr.202100013}
\end{barticle}
\endbibitem

\bibitem[\protect\citeauthoryear{Wu et~al.}{2017}]{WuX2017}
\begin{barticle}
\bauthor{\bsnm{Wu}, \binits{X.}},
\bauthor{\bsnm{Meng}, \binits{Y.}},
\bauthor{\bsnm{Tian}, \binits{J.}},
\bauthor{\bsnm{Huang}, \binits{Y.}},
\bauthor{\bsnm{Xiang}, \binits{H.}},
\bauthor{\bsnm{Han}, \binits{D.}},
\bauthor{\bsnm{Wen}, \binits{W.}}:
\batitle{Direct observation of valley-polarized topological edge states in designer surface plasmon crystals}.
\bjtitle{Nature Communications}
\bvolume{8}(\bissue{1}),
\bfpage{1304}
(\byear{2017})
\doiurl{10.1038/s41467-017-01515-2}
\end{barticle}
\endbibitem

\bibitem[\protect\citeauthoryear{He et~al.}{2019}]{HeX2019}
\begin{barticle}
\bauthor{\bsnm{He}, \binits{X.-T.}},
\bauthor{\bsnm{Liang}, \binits{E.-T.}},
\bauthor{\bsnm{Yuan}, \binits{J.-J.}},
\bauthor{\bsnm{Qiu}, \binits{H.-Y.}},
\bauthor{\bsnm{Chen}, \binits{X.-D.}},
\bauthor{\bsnm{Zhao}, \binits{F.-L.}},
\bauthor{\bsnm{Dong}, \binits{J.-W.}}:
\batitle{A silicon-on-insulator slab for topological valley transport}.
\bjtitle{Nature Communications}
\bvolume{10}(\bissue{1}),
\bfpage{872}
(\byear{2019})
\doiurl{10.1038/s41467-019-08881-z}
\end{barticle}
\endbibitem

\bibitem[\protect\citeauthoryear{Yoshimi et~al.}{2020}]{Hironobu2020}
\begin{barticle}
\bauthor{\bsnm{Yoshimi}, \binits{H.}},
\bauthor{\bsnm{Yamaguchi}, \binits{T.}},
\bauthor{\bsnm{Ota}, \binits{Y.}},
\bauthor{\bsnm{Arakawa}, \binits{Y.}},
\bauthor{\bsnm{Iwamoto}, \binits{S.}}:
\batitle{Slow light waveguides in topological valley photonic crystals}.
\bjtitle{Optics Letters}
\bvolume{45}(\bissue{9}),
\bfpage{2648}--\blpage{2651}
(\byear{2020})
\doiurl{10.1364/OL.391764}
\end{barticle}
\endbibitem

\bibitem[\protect\citeauthoryear{Kumar et~al.}{2024}]{Kumar2024}
\begin{barticle}
\bauthor{\bsnm{Kumar}, \binits{A.}},
\bauthor{\bsnm{Tan}, \binits{Y.J.}},
\bauthor{\bsnm{Navaratna}, \binits{N.}},
\bauthor{\bsnm{Gupta}, \binits{M.}},
\bauthor{\bsnm{Pitchappa}, \binits{P.}},
\bauthor{\bsnm{Singh}, \binits{R.}}:
\batitle{Slow light topological photonics with counter-propagating waves and its active control on a chip}.
\bjtitle{Nature Communications}
\bvolume{15}(\bissue{1}),
\bfpage{926}
(\byear{2024})
\doiurl{10.1038/s41467-024-45175-5}
\end{barticle}
\endbibitem

\bibitem[\protect\citeauthoryear{Iwamoto et~al.}{2021}]{Iwamoto2021}
\begin{barticle}
\bauthor{\bsnm{Iwamoto}, \binits{S.}},
\bauthor{\bsnm{Ota}, \binits{Y.}},
\bauthor{\bsnm{Arakawa}, \binits{Y.}}:
\batitle{Recent progress in topological waveguides and nanocavities in a semiconductor photonic crystal platform}.
\bjtitle{Optical Materials Express}
\bvolume{11}(\bissue{2}),
\bfpage{319}--\blpage{337}
(\byear{2021})
\doiurl{10.1364/OME.415128}
\end{barticle}
\endbibitem

\bibitem[\protect\citeauthoryear{Lan et~al.}{2022}]{Zhihao2022}
\begin{barticle}
\bauthor{\bsnm{Lan}, \binits{Z.}},
\bauthor{\bsnm{Chen}, \binits{M.L.N.}},
\bauthor{\bsnm{Gao}, \binits{F.}},
\bauthor{\bsnm{Zhang}, \binits{S.}},
\bauthor{\bsnm{Sha}, \binits{W.E.I.}}:
\batitle{A brief review of topological photonics in one, two, and three dimensions}.
\bjtitle{Reviews in Physics}
\bvolume{9},
\bfpage{100076}
(\byear{2022})
\doiurl{10.1016/j.revip.2022.100076}
\end{barticle}
\endbibitem

\bibitem[\protect\citeauthoryear{Ozawa et~al.}{2019}]{Ozawa2019}
\begin{barticle}
\bauthor{\bsnm{Ozawa}, \binits{T.}},
\bauthor{\bsnm{Price}, \binits{H.M.}},
\bauthor{\bsnm{Amo}, \binits{A.}},
\bauthor{\bsnm{Goldman}, \binits{N.}},
\bauthor{\bsnm{Hafezi}, \binits{M.}},
\bauthor{\bsnm{Lu}, \binits{L.}},
\bauthor{\bsnm{Rechtsman}, \binits{M.C.}},
\bauthor{\bsnm{Schuster}, \binits{D.}},
\bauthor{\bsnm{Simon}, \binits{J.}},
\bauthor{\bsnm{Zilberberg}, \binits{O.}},
\bauthor{\bsnm{Carusotto}, \binits{I.}}:
\batitle{Topological photonics}.
\bjtitle{Reviews of Modern Physics}
\bvolume{91},
\bfpage{015006}
(\byear{2019})
\doiurl{10.1103/RevModPhys.91.015006}
\end{barticle}
\endbibitem

\bibitem[\protect\citeauthoryear{Muis et~al.}{2025}]{Muis2025}
\begin{barticle}
\bauthor{\bsnm{Muis}, \binits{D.}},
\bauthor{\bsnm{Li}, \binits{Y.}},
\bauthor{\bsnm{Barczyk}, \binits{R.}},
\bauthor{\bsnm{Arora}, \binits{S.}},
\bauthor{\bsnm{Kuipers}, \binits{L.}},
\bauthor{\bsnm{Shvets}, \binits{G.}},
\bauthor{\bsnm{Verhagen}, \binits{E.}}:
\batitle{Broadband localization of light at the termination of a topological photonic waveguide}.
\bjtitle{Science Advances}
\bvolume{11}(\bissue{16}),
\bfpage{9569}
(\byear{2025})
\doiurl{10.1126/sciadv.adr9569}
\end{barticle}
\endbibitem

\bibitem[\protect\citeauthoryear{Zhang et~al.}{2018}]{ZhangX2018}
\begin{barticle}
\bauthor{\bsnm{Zhang}, \binits{X.}},
\bauthor{\bsnm{Xiao}, \binits{M.}},
\bauthor{\bsnm{Cheng}, \binits{Y.}},
\bauthor{\bsnm{Lu}, \binits{M.-H.}},
\bauthor{\bsnm{Christensen}, \binits{J.}}:
\batitle{Topological sound}.
\bjtitle{Communications Physics}
\bvolume{1}(\bissue{1}),
\bfpage{97}
(\byear{2018})
\doiurl{10.1038/s42005-018-0094-4}
\end{barticle}
\endbibitem

\bibitem[\protect\citeauthoryear{Wang et~al.}{2022}]{WangJ2022}
\begin{barticle}
\bauthor{\bsnm{Wang}, \binits{J.-Q.}},
\bauthor{\bsnm{Zhang}, \binits{Z.-D.}},
\bauthor{\bsnm{Yu}, \binits{S.-Y.}},
\bauthor{\bsnm{Ge}, \binits{H.}},
\bauthor{\bsnm{Liu}, \binits{K.-F.}},
\bauthor{\bsnm{Wu}, \binits{T.}},
\bauthor{\bsnm{Sun}, \binits{X.-C.}},
\bauthor{\bsnm{Liu}, \binits{L.}},
\bauthor{\bsnm{Chen}, \binits{H.-Y.}},
\bauthor{\bsnm{He}, \binits{C.}},
\bauthor{\bsnm{Lu}, \binits{M.-H.}},
\bauthor{\bsnm{Chen}, \binits{Y.-F.}}:
\batitle{Extended topological valley-locked surface acoustic waves}.
\bjtitle{Nature Communications}
\bvolume{13}(\bissue{1}),
\bfpage{1324}
(\byear{2022})
\doiurl{10.1038/s41467-022-29019-8}
\end{barticle}
\endbibitem

\bibitem[\protect\citeauthoryear{Zhang et~al.}{2021}]{ZhangZ2021}
\begin{barticle}
\bauthor{\bsnm{Zhang}, \binits{Z.-D.}},
\bauthor{\bsnm{Yu}, \binits{S.-Y.}},
\bauthor{\bsnm{Ge}, \binits{H.}},
\bauthor{\bsnm{Wang}, \binits{J.-Q.}},
\bauthor{\bsnm{Wang}, \binits{H.-F.}},
\bauthor{\bsnm{Liu}, \binits{K.-F.}},
\bauthor{\bsnm{Wu}, \binits{T.}},
\bauthor{\bsnm{He}, \binits{C.}},
\bauthor{\bsnm{Lu}, \binits{M.-H.}},
\bauthor{\bsnm{Chen}, \binits{Y.-F.}}:
\batitle{Topological surface acoustic waves}.
\bjtitle{Physical Review Applied}
\bvolume{16},
\bfpage{044008}
(\byear{2021})
\doiurl{10.1103/PhysRevApplied.16.044008}
\end{barticle}
\endbibitem

\bibitem[\protect\citeauthoryear{Zhou et~al.}{2023}]{ZhouY2023}
\begin{barticle}
\bauthor{\bsnm{Zhou}, \binits{Y.}},
\bauthor{\bsnm{Zhang}, \binits{N.}},
\bauthor{\bsnm{Bisharat}, \binits{D.J.}},
\bauthor{\bsnm{Davis}, \binits{R.J.}},
\bauthor{\bsnm{Zhang}, \binits{Z.}},
\bauthor{\bsnm{Friend}, \binits{J.}},
\bauthor{\bsnm{Bandaru}, \binits{P.R.}},
\bauthor{\bsnm{Sievenpiper}, \binits{D.F.}}:
\batitle{On-chip unidirectional waveguiding for surface acoustic waves along a defect line in a triangular lattice}.
\bjtitle{Physical Review Applied}
\bvolume{19},
\bfpage{024053}
(\byear{2023})
\doiurl{10.1103/PhysRevApplied.19.024053}
\end{barticle}
\endbibitem

\bibitem[\protect\citeauthoryear{Yang et~al.}{2015}]{YangZ2015}
\begin{barticle}
\bauthor{\bsnm{Yang}, \binits{Z.}},
\bauthor{\bsnm{Gao}, \binits{F.}},
\bauthor{\bsnm{Shi}, \binits{X.}},
\bauthor{\bsnm{Lin}, \binits{X.}},
\bauthor{\bsnm{Gao}, \binits{Z.}},
\bauthor{\bsnm{Chong}, \binits{Y.}},
\bauthor{\bsnm{Zhang}, \binits{B.}}:
\batitle{Topological acoustics}.
\bjtitle{Physical Review Letters}
\bvolume{114},
\bfpage{114301}
(\byear{2015})
\doiurl{10.1103/PhysRevLett.114.114301}
\end{barticle}
\endbibitem

\bibitem[\protect\citeauthoryear{Ma et~al.}{2019}]{MaG2019}
\begin{barticle}
\bauthor{\bsnm{Ma}, \binits{G.}},
\bauthor{\bsnm{Xiao}, \binits{M.}},
\bauthor{\bsnm{Chan}, \binits{C.T.}}:
\batitle{Topological phases in acoustic and mechanical systems}.
\bjtitle{Nature Reviews Physics}
\bvolume{1}(\bissue{1}),
\bfpage{281}--\blpage{294}
(\byear{2019})
\doiurl{10.1038/s42254-019-0030-x}
\end{barticle}
\endbibitem

\bibitem[\protect\citeauthoryear{Ma et~al.}{2021}]{MaJ2021}
\begin{barticle}
\bauthor{\bsnm{Ma}, \binits{J.}},
\bauthor{\bsnm{Xi}, \binits{X.}},
\bauthor{\bsnm{Sun}, \binits{X.}}:
\batitle{Experimental demonstration of dual-band nano-electromechanical valley-hall topological metamaterials}.
\bjtitle{Advanced Materials}
\bvolume{33}(\bissue{10}),
\bfpage{2006521}
(\byear{2021})
\doiurl{10.1002/adma.202006521}
\end{barticle}
\endbibitem

\bibitem[\protect\citeauthoryear{Funayama et~al.}{2025}]{Funayama2025}
\begin{barticle}
\bauthor{\bsnm{Funayama}, \binits{K.}},
\bauthor{\bsnm{Yatsugi}, \binits{K.}},
\bauthor{\bsnm{Tanaka}, \binits{H.}},
\bauthor{\bsnm{Iizuka}, \binits{H.}}:
\batitle{Quantum valley hall effect-based coupler with continuously tunable transmission for topological information communication}.
\bjtitle{Advanced Science}
\bvolume{12}(\bissue{35}),
\bfpage{06732}
(\byear{2025})
\doiurl{10.1002/advs.202506732}
\end{barticle}
\endbibitem

\bibitem[\protect\citeauthoryear{Yan et~al.}{2018}]{YanM2018}
\begin{barticle}
\bauthor{\bsnm{Yan}, \binits{M.}},
\bauthor{\bsnm{Lu}, \binits{J.}},
\bauthor{\bsnm{Li}, \binits{F.}},
\bauthor{\bsnm{Deng}, \binits{W.}},
\bauthor{\bsnm{Huang}, \binits{X.}},
\bauthor{\bsnm{Ma}, \binits{J.}},
\bauthor{\bsnm{Liu}, \binits{Z.}}:
\batitle{On-chip valley topological materials for elastic wave manipulation}.
\bjtitle{Nature Materials}
\bvolume{17}(\bissue{11}),
\bfpage{993}--\blpage{998}
(\byear{2018})
\doiurl{10.1038/s41563-018-0191-5}
\end{barticle}
\endbibitem

\bibitem[\protect\citeauthoryear{Darabi et~al.}{2020}]{Darabi2020}
\begin{barticle}
\bauthor{\bsnm{Darabi}, \binits{A.}},
\bauthor{\bsnm{Ni}, \binits{X.}},
\bauthor{\bsnm{Leamy}, \binits{M.}},
\bauthor{\bsnm{Alù}, \binits{A.}}:
\batitle{Reconfigurable floquet elastodynamic topological insulator based on synthetic angular momentum bias}.
\bjtitle{Science Advances}
\bvolume{6}(\bissue{29}),
\bfpage{8656}
(\byear{2020})
\doiurl{10.1126/sciadv.aba8656}
\end{barticle}
\endbibitem

\bibitem[\protect\citeauthoryear{Funayama et~al.}{2024}]{Funayama202412}
\begin{barticle}
\bauthor{\bsnm{Funayama}, \binits{K.}},
\bauthor{\bsnm{Yatsugi}, \binits{K.}},
\bauthor{\bsnm{Iizuka}, \binits{H.}}:
\batitle{Quantum valley hall effect-based topological boundaries for frequency-dependent and -independent mode energy profiles}.
\bjtitle{Communications Physics}
\bvolume{7}(\bissue{1}),
\bfpage{409}
(\byear{2024})
\doiurl{10.1038/s42005-024-01899-w}
\end{barticle}
\endbibitem

\bibitem[\protect\citeauthoryear{Monika~Devi et~al.}{2021}]{MonikaDevi2021}
\begin{barticle}
\bauthor{\bsnm{Monika~Devi}, \binits{K.}},
\bauthor{\bsnm{Jana}, \binits{S.}},
\bauthor{\bsnm{Roy~Chowdhury}, \binits{D.}}:
\batitle{Topological edge states in an all-dielectric terahertz photonic crystal}.
\bjtitle{Optical Materials Express}
\bvolume{11}(\bissue{8}),
\bfpage{2445}--\blpage{2458}
(\byear{2021})
\doiurl{10.1364/OME.427069}
\end{barticle}
\endbibitem

\bibitem[\protect\citeauthoryear{Wong et~al.}{2020}]{WongS2020}
\begin{barticle}
\bauthor{\bsnm{Wong}, \binits{S.}},
\bauthor{\bsnm{Saba}, \binits{M.}},
\bauthor{\bsnm{Hess}, \binits{O.}},
\bauthor{\bsnm{Oh}, \binits{S.S.}}:
\batitle{Gapless unidirectional photonic transport using all-dielectric kagome lattices}.
\bjtitle{Physical Review Research}
\bvolume{2},
\bfpage{012011}
(\byear{2020})
\doiurl{10.1103/PhysRevResearch.2.012011}
\end{barticle}
\endbibitem

\bibitem[\protect\citeauthoryear{Vakulenko et~al.}{2023}]{Vakulenko2023}
\begin{barticle}
\bauthor{\bsnm{Vakulenko}, \binits{A.}},
\bauthor{\bsnm{Kiriushechkina}, \binits{S.}},
\bauthor{\bsnm{Smirnova}, \binits{D.}},
\bauthor{\bsnm{Guddala}, \binits{S.}},
\bauthor{\bsnm{Komissarenko}, \binits{F.}},
\bauthor{\bsnm{Alù}, \binits{A.}},
\bauthor{\bsnm{Allen}, \binits{M.}},
\bauthor{\bsnm{Allen}, \binits{J.}},
\bauthor{\bsnm{Khanikaev}, \binits{A.B.}}:
\batitle{Adiabatic topological photonic interfaces}.
\bjtitle{Nature Communications}
\bvolume{14}(\bissue{1}),
\bfpage{4629}
(\byear{2023})
\doiurl{10.1038/s41467-023-40238-5}
\end{barticle}
\endbibitem

\bibitem[\protect\citeauthoryear{Shah et~al.}{2021}]{Shah2021}
\begin{barticle}
\bauthor{\bsnm{Shah}, \binits{T.}},
\bauthor{\bsnm{Marquardt}, \binits{F.}},
\bauthor{\bsnm{Peano}, \binits{V.}}:
\batitle{Tunneling in the brillouin zone: Theory of backscattering in valley hall edge channels}.
\bjtitle{Physical Review B}
\bvolume{104},
\bfpage{235431}
(\byear{2021})
\doiurl{10.1103/PhysRevB.104.235431}
\end{barticle}
\endbibitem

\bibitem[\protect\citeauthoryear{Funayama et~al.}{2022}]{Funayama2021}
\begin{barticle}
\bauthor{\bsnm{Funayama}, \binits{K.}},
\bauthor{\bsnm{Yatsugi}, \binits{K.}},
\bauthor{\bsnm{Miura}, \binits{A.}},
\bauthor{\bsnm{Iizuka}, \binits{H.}}:
\batitle{Control of coupling between micromechanical topological waveguides}.
\bjtitle{International Journal of Mechanical Sciences}
\bvolume{236},
\bfpage{107755}
(\byear{2022})
\doiurl{10.1016/j.ijmecsci.2022.107755}
\end{barticle}
\endbibitem

\bibitem[\protect\citeauthoryear{Tian et~al.}{2020}]{Tian2020}
\begin{barticle}
\bauthor{\bsnm{Tian}, \binits{Z.}},
\bauthor{\bsnm{Shen}, \binits{C.}},
\bauthor{\bsnm{Li}, \binits{J.}},
\bauthor{\bsnm{Reit}, \binits{E.}},
\bauthor{\bsnm{Bachman}, \binits{H.}},
\bauthor{\bsnm{Socolar}, \binits{J.E.S.}},
\bauthor{\bsnm{Cummer}, \binits{S.A.}},
\bauthor{\bsnm{Jun~Huang}, \binits{T.}}:
\batitle{Dispersion tuning and route reconfiguration of acoustic waves in valley topological phononic crystals}.
\bjtitle{Nature Communications}
\bvolume{11}(\bissue{1}),
\bfpage{762}
(\byear{2020})
\doiurl{10.1038/s41467-020-14553-0}
\end{barticle}
\endbibitem

\bibitem[\protect\citeauthoryear{Chen et~al.}{2022}]{ChenL2022}
\begin{barticle}
\bauthor{\bsnm{Chen}, \binits{L.}},
\bauthor{\bsnm{Zhao}, \binits{M.}},
\bauthor{\bsnm{Ye}, \binits{H.}},
\bauthor{\bsnm{Hang}, \binits{Z.H.}},
\bauthor{\bsnm{Li}, \binits{Y.}},
\bauthor{\bsnm{Cao}, \binits{Z.}}:
\batitle{Efficient light coupling between conventional silicon photonic waveguides and quantum valley-hall topological interfaces}.
\bjtitle{Optics Express}
\bvolume{30}(\bissue{2}),
\bfpage{2517}--\blpage{2527}
(\byear{2022})
\doiurl{10.1364/OE.445851}
\end{barticle}
\endbibitem

\bibitem[\protect\citeauthoryear{Huo et~al.}{2017}]{HuoS2017}
\begin{barticle}
\bauthor{\bsnm{Huo}, \binits{S.-y.}},
\bauthor{\bsnm{Chen}, \binits{J.-j.}},
\bauthor{\bsnm{Huang}, \binits{H.-b.}},
\bauthor{\bsnm{Huang}, \binits{G.-l.}}:
\batitle{Simultaneous multi-band valley-protected topological edge states of shear vertical wave in two-dimensional phononic crystals with veins}.
\bjtitle{Scientific Reports}
\bvolume{7}(\bissue{1}),
\bfpage{10335}
(\byear{2017})
\doiurl{10.1038/s41598-017-10857-2}
\end{barticle}
\endbibitem

\end{thebibliography}

\end{document}